\documentclass[%
preprintm, superscriptaddress, 12pt,
 amsmath,amssymb,
 aps,
]{revtex4-2}

\usepackage[utf8]{inputenc}
\usepackage{scalerel}
\usepackage{siunitx}
\usepackage{dcolumn} 
\usepackage{titlesec}
\titlespacing*{\section}
{0pt}{4.5ex plus 1ex minus .2ex}{1.3ex plus .2ex}
\titleformat{\section}
  {\centering\bfseries\MakeUppercase}{\thesection}{1em}{}
\titleformat{\subsection}
  {\centering\bfseries\MakeUppercase}{\thesubsection}{1em}{}
  \titleformat{\subsubsection}
  {\centering\bfseries\MakeUppercase}{\thesubsubsection}{1em}{}
  
\usepackage{xstring}
\usepackage{tikz}

\usepackage{fancyhdr} 
\usepackage{titlesec}
\titlespacing*{\section}
{0pt}{4.5ex plus 1ex minus .2ex}{1.3ex plus .2ex}
\titleformat{\section}
  {\centering\bfseries\MakeUppercase}{\thesection}{1em}{}

\usepackage{graphicx}
\usepackage{dcolumn}
\usepackage{float}

\usepackage{mathtools}
\usepackage{bm}
\usepackage{gensymb} 

\usepackage{natbib}
\usepackage[english]{babel}
\usepackage[utf8]{inputenc}
\usepackage{url}
\usepackage[colorlinks = true,
            linkcolor = blue,
            urlcolor  = blue,
            citecolor = blue,
            anchorcolor = blue]{hyperref}

\begin{document}

\title{Tutorial: Characterizing Microscale Energy Transport in Materials with Transient Grating Spectroscopy}

\author{Usama Choudhry}
\affiliation{Department of Mechanical Engineering, University of California, Santa Barbara, CA 93106, USA}

\author{Taeyong Kim}
\affiliation{Department of Mechanical Engineering, University of California, Santa Barbara, CA 93106, USA}

\author{Melanie Adams}
\affiliation{Department of Mechanical Engineering, University of California, Santa Barbara, CA 93106, USA}

\author{Jeewan Ranasinghe}
\affiliation{Department of Mechanical Engineering, University of California, Santa Barbara, CA 93106, USA}

\author{Runqing Yang}
\affiliation{Department of Mechanical Engineering, University of California, Santa Barbara, CA 93106, USA}

\author{Bolin Liao}
\email{bliao@ucsb.edu} 
\affiliation{Department of Mechanical Engineering, University of California, Santa Barbara, CA 93106, USA}

\date{\today}

\begin{abstract}
Microscale energy transport processes are crucial in microelectronics, energy harvesting devices and emerging quantum materials. To study these processes, methods that can probe transport with conveniently tunable length scales are highly desirable. Transient grating spectroscopy (TGS) is such a tool that can monitor microscale energy transport processes associated with various fundamental energy carriers including electrons, phonons and spins. Having been developed and applied for a long time in the chemistry community, TGS has regained popularity recently in studying different transport regimes in solid-state materials. In this Tutorial, we provide an in-depth discussion of the operational principle and instrumentation details of a modern heterodyne TGS configuration from a practitioner's point of view. We further review recent applications of TGS in characterizing microscale transport of heat, charge, spin and acoustic waves, with an emphasis on thermal transport. 

\end{abstract}

\maketitle
\section{Introduction}
Fundamental understanding of energy transport processes on the micro- and nano-scale is essential for a broad range of applications including electronics, opto-electronics, and energy harvesting devices~\cite{cahill2014nanoscale,liao2020nanoscale}. In particular, as the devices' characteristic time and length scales become comparable to or even smaller than the intrinsic time and length scales associated with the fundamental energy carriers, new transport regimes, such as ballistic and coherent transport, will emerge, where the transport properties can deviate qualitatively from the predictions of classical continuum laws~\cite{chen2021non}. For this reason, intense research efforts have focused on developing tools to probe energy transport with increasing time and space resolutions. For example, variations of ultrafast optical spectroscopy, such as time-domain thermoreflectance (TDTR)~\cite{jiang2018tutorial} and transient absorption microscopy (TAM)~\cite{guo2017long}, have been applied to measure microscale thermal transport and photocarrier transport, respectively; the recent development of ultrafast electron diffraction\cite{yang2020simultaneous} and microscopy~\cite{zewail2010four,liao2017scanning} has opened new doors towards time-resolved imaging of energy transport in sub-picosecond and nanometer scale.

Among these techniques, transient grating spectroscopy (TGS) is a versatile tool that can detect the transport and relaxation processes of a broad range of energy carriers, such as electrons, phonons and spins, across multiple time and length scales~\cite{eichler1986laser}. Also called impulsive stimulated light scattering (ISS), TGS utilizes a pair of optical pump beams that cross and coherently superpose at the sample location. The interference of the two pump beams forms a one-dimensional (1D) periodic spatial grating that leads to a 1D periodic material response. The temporal decay of this periodic material response, or the ``transient grating'', encodes information about local relaxation and transport, and can be detected by an optical probe beam that is diffracted off the transient grating. Originally developed in the 1970s and 80s mainly in the chemistry community~\cite{eichler1986laser}, TGS has gained increasing interest for applications in materials science and thermal engineering. Recent breakthroughs utilizing TGS include the measurement of quasi-particle diffusion in a high-temperature superconductor~\cite{gedik2003diffusion}, the discovery of spin-Coulomb drag in a GaAs quantum well~\cite{weber2005observation}, quasi-ballistic phonon transport in a silicon membrane at room temperature~\cite{johnson2013direct}, and the detection of second sound in graphite above 100 $K$~\cite{huberman2019observation}. 

Compared to the other techniques mentioned above, TGS has the following advantages: (i) Flexibility. The transport length scale, which is the spatial period of the transient grating, can be conveniently tuned in TGS by varying the crossing angle of the pump beams. By simply adjusting the polarizations of the two pump beams, either an intensity grating or a polarization grating~\cite{mahmood2018observation} can be formed that can couple to different degrees of freedom in a sample. Configurations with time resolutions ranging from nanosecond to femtosecond~\cite{kim2017elastic} have been demonstrated. (ii) Sensitivity to in-plane transport. Due to the transient grating forming along the in-plane direction, TGS is naturally more sensitive to in-plane transport. For this reason, TGS has been combined with other techniques that are more sensitive to cross-plane transport, such as TDTR, to characterize anisotropic materials~\cite{luckyanova2013anisotropy,li2021remarkably}. (iii) Convenient implementation of heterodyne detection~\cite{maznev1998optical}. As will be discussed in detail in Section \ref{sec:heterodyne_detection}, heterodyne detection of transient gratings not only boosts the signal level and cancels unwanted background, but also allows for the measurement of separate contributions from light-induced changes in the real and imaginary parts of the complex refractive index. (iv) Facile sample preparation. Compared to some other techniques, such as TDTR, which typically requires the deposition of a thin metal transducer layer, TGS does not require further processing of the sample as long as the sample surfaces are optically smooth. (v) Facilitated theoretical analysis. Owing to the sinusoidal excitation profile, the TGS signal can be modeled by including only one Fourier component of the material response, which often simplifies the theoretical analysis.

In this Tutorial, we will discuss the instrumentation details and operational principles of a modern heterodyne TGS configuration~\cite{maznev1998optical} from a practitioner's point of view. Then we will review recent advancements in exploiting TGS to study various forms of microscale energy transport, with a particular focus on thermal transport. Most of the materials about TGS setup and operation covered here can be found in the classic text~\cite{eichler1986laser}, excellent papers~\cite{kading1995transient,maznev1998optical,johnson2012phase,hofmann2019transient} and theses~\cite{johnson2011optical,robbins2019exploring,kim2020investigation}. The purpose of this Tutorial is to distill the essential information from previous literature and, together with our own experience of operating a TGS setup, provide a concise guide for workers who are interested in using TGS 
for their own research. This Tutorial is organized as follows. In Section \ref{sec:operational_principles}, we will elaborate the basic operational principles of TGS, including the formation of optical gratings, the material response (including a detailed discussion on thermal models), and the grating detection focusing on the optical heterodyne method. In Section \ref{sec:thermal_transport_applications}, we review recent applications of TGS on characterizing nanoscale thermal transport, with a focus on phonon mean free path spectroscopy~\cite{johnson2013direct,robbins2019ballistic} and second sound detection~\cite{huberman2019observation}. In Section \ref{sec:other_applications}, we review recent examples utilizing TGS to study photocarrier, spin, and surface acoustic wave (SAW) dynamics. We summarize the Tutorial and share our perspectives on future developments of TGS in Section \ref{sec:summary}.  

\section{Operational Principles and Instrumentation}
\label{sec:operational_principles}
\begin{figure}[hbt!]
\centering
\includegraphics[width=\columnwidth,keepaspectratio]{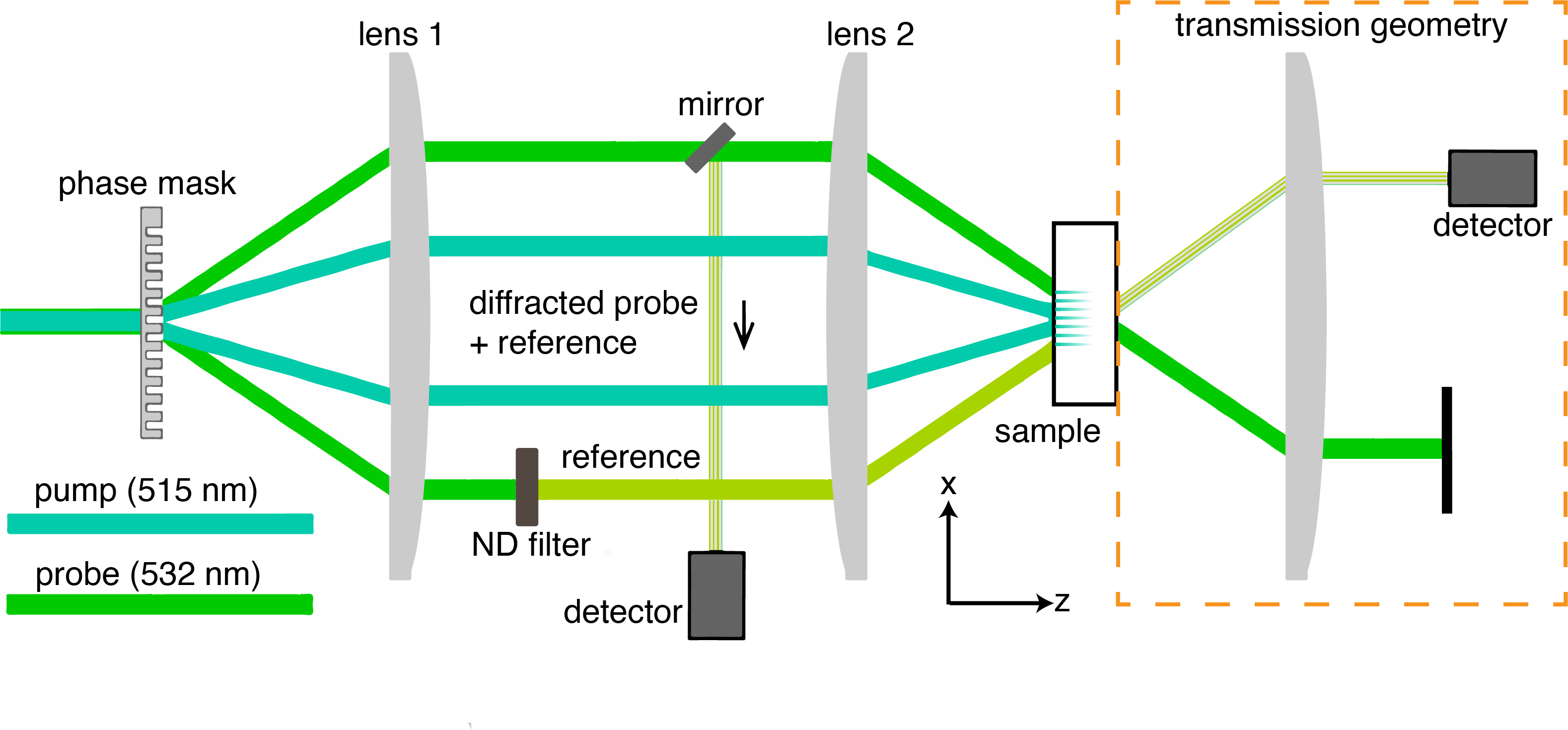}
\caption{\textbf{Schematic Illustration of A Typical TGS Experiment.} The schematic is for a reflection geometry. The components inside the dashed box is for a transmission geometry.}
\label{fig:schematic}
\end{figure}

Earlier implementations of TGS used separate optical paths to control the two pump beams, the probe beam and the reference beam (for heterodyne detection)~\cite{eichler1986laser}. The alignment in these implementations was challenging and the phase instability of the beams could be severe enough to disrupt the heterodyne detection, which necessitated active phase stabilization in some cases~\cite{eichler1986laser,matsuo1997phase}. To overcome these practical challenges, Maznev et al.\cite{maznev1998optical} proposed a compact TGS setup utilizing a phase mask (a transparent diffractive grating), as illustrated in Fig.~\ref{fig:schematic}, which has been broadly adopted in more recent TGS implementations. In this Tutorial, we will focus on this setup. We analyze the operational principle of this TGS setup in three steps: grating formation, materials response, and grating detection.    

\subsection{Grating Formation}
\label{sec:grating_formation}
As shown in Fig.~\ref{fig:schematic}, the incoming pump beam with wavelength $\lambda_{pp}$ and probe beam with wavelength $\lambda_{pr}$ are diffracted by a phase mask with a spatial period of $L_{PM}$. For TGS setups with nanosecond time resolution, typically the pump beam is generated from a pulsed laser with nanosecond pulse width, while the probe beam is from a continuous-wave (CW) laser (often modulated by a mechanical chopper or an optical modulator to reduce steady sample heating). For femtosecond time resolution TGS setups, both pump and probe beams are from a femtosecond pulsed laser. The phase mask is typically designed to optimize the diffraction efficiency for the $\pm 1$-order beams. The angle between the two diffracted beams of $\pm 1$ order is determined by the phase mask period $L_{PM}$ and the light wavelengths:
\begin{equation}
    \theta_{1,pp/pr}=2\arcsin{\frac{\lambda_{pp/pr}}{L_{PM}}} \approx \frac{2\lambda_{pp/pr}}{L_{PM}},
\end{equation}
assuming $\lambda_{pp/pr} \ll L_{PM}$ so the beams remain close to the optical axis. The diffracted beams are collimated and refocused by two lenses with focal lengths $f_1$ and $f_2$, respectively, in a telescope configuration onto the sample surface. To maximize the overlap of the pump and the probe beams on the sample surface, chromatic aberration of the two lenses should be minimized such that the focal lengths remain the same at the pump and the probe wavelengths. Typically, achromatic doublets are used for this purpose and the pump/probe wavelengths are chosen to be close to each other. For example, a commonly used pump/probe pair is 515 nm/532 nm. The converging angle of the two pump/probe beams is determined by the focal lengths of the two lenses and given, assuming near-axial condition, as
\begin{equation}
    \theta_{2,pp/pr} \approx \frac{f_1}{f_2}\theta_{1,pp/pr} = 2 \frac{f_1}{f_2} \frac{\lambda_{pp/pr}}{L_{PM}}.
    \label{eqn:pump_probe_angle}
\end{equation}

Assuming the sample has a refractive index of $n$, the wavelengths of the pump/probe beams become $\lambda_{pp/pr}/n$ inside the sample. The electric fields associated with the pump beams can be described as:
\begin{equation}
    \bm{E}_{\rm{pp}\pm}=E_{pp} \hat{\bm{p}}_{\pm} e^{-i (\omega_{pp} t - k_{z,pp} z \pm k_{x,pp} x)},
\end{equation}
where $E_{pp}$ is the magnitude of the field, $\omega_{pp}$ is the angular frequency of the pump light, $\bm{p}_{\pm}$ is the polarization of the $\pm 1$-order beam, $k_{x,pp} = \frac{2 \pi n}{\lambda_{pp}} \sin{\frac{\theta_{3,pp}}{2}}$, and $k_{z,pp} = \frac{2 \pi n}{\lambda_{pp}} \cos{\frac{\theta_{3,pp}}{2}}$. $\theta_{3,pp/pr}$ is the converging angle of the pump/probe beams inside the sample and connected to $\theta_{2,pp/pr}$ via Snell's law: $n\sin{\theta_{3,pp/pr}}=\sin{\theta_{2,pp/pr}}$. 
The superposed electric field in the interference region is $\bm{E}_{\rm{pp}} = \bm{E}_{\rm{pp}+}+\bm{E}_{\rm{pp}-}$. The distribution of the intensity of the pump light in the interference region is:
\begin{equation}
    I_{\rm{pp}}(x) \propto |\bm{E}_{\rm{pp}}|^2 = 2 {E_{pp}}^2 [1+\hat{\bm{p}}_{+} \cdot \hat{\bm{p}}_{-} \cos{2 k_{x,pp} x}].
\end{equation}

When the polarization of the two pump beams are parallel to each other, the intensity grating strength is maximized with the grating period $L_S$ inside the sample:
\begin{equation}
    L_S=\frac{2\pi}{2 k_{x,pp}}=\frac{\lambda_{pp}}{2 n \sin{\frac{\theta_{3,pp}}{2}}} = \frac{\lambda_{pp}}{2 \sin{\frac{\theta_{2,pp}}{2}}} \approx \frac{f_2}{2 f_1} L_{PM}.
    \label{eqn:grating_period}
\end{equation}
We can also define $q = \frac{2 \pi}{L_S}$ as the wavevector associated with the optical grating. From Eqn.~\ref{eqn:grating_period}, the period of the optical grating is determined by the period of the phase mask and the focal lengths of the lenses, which essentially image the phase mask pattern onto the sample surface. A typical phase mask consists of a transparent (e.g. glass) substrate with a set of grating patterns with different periods etched on the substrate. So the period of the optical grating formed on the sample surface can be easily adjusted by translating the phase mask between different etched grating patterns\cite{maznev2003laser}. In addition, continuous change of the optical grating period can be attained by slightly tilting the phase mask\cite{vega2015laser}. In practice, due to nonidealities in the optical system, the actual grating period on the sample may deviate from Eqn.~\ref{eqn:grating_period}. Therefore, the actual grating period should be calibrated, e.g. by burning a physical grating on a calibration sample or measuring acoustic oscillation frequencies in a sample with known acoustic properties\cite{kim2020investigation}. While the theoretical minimum of the grating period is determined by the diffraction limit $L_{min} = \frac{\lambda_{pp}}{2n}$, practically achievable minimum grating period is further limited by the numerical aperture of the optical system. Using four-inch optics, Robbins et al. demonstrated an optical grating period of 577 nm using a 515-nm pump laser\cite{robbins2019ballistic}. Another approach to obtain very small grating periods is to reduce the pump wavelength, for example with extreme ultraviolet (EUV) sources\cite{bencivenga2019nanoscale}. On the other hand, the largest grating period achievable is often limited by the pump beam size and the small spatial separation between the beams that can create challenges in arranging the optics. Typically, grating periods up to tens of micrometers are used in TGS experiments while the pump beam size is on the order of a few hundred micrometers\cite{johnson2011optical,robbins2019ballistic}.

\begin{figure}[hbt!]
\centering
\includegraphics[width=\columnwidth,keepaspectratio]{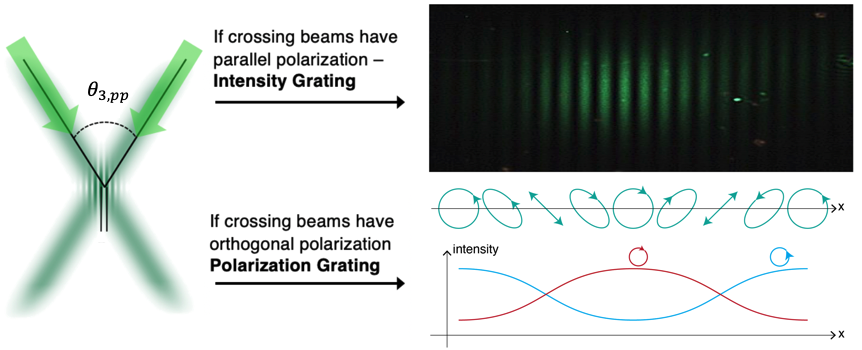}
\caption{\textbf{Schematic Illustration of Intensity Grating and Polarization Grating.} When the polarizations of the two pump beams are parallel, an intensity grating is formed. When the polarizations of the two pump beams are perpendicular, a polarization grating is formed. The polarization grating can be decomposed into two intensity gratings with opposite senses of circular polarization and a $\pi$ spatial phase shift.}
\label{fig:grating}
\end{figure}

If the two pump beams have perpendicular polarizations, the resulting intensity distribution is uniform, but the polarization will change periodically. For example, if the two pump beams are polarized along the $\hat{\bm{x}}$ and $\hat{\bm{y}}$ directions, respectively, the resulting superposed electric field is:
\begin{equation}
    \bm{E}_{\rm{pp}} = E_{pp} e^{-i(\omega_{pp} t - k_{z,pp} z)} (\hat{\bm{x}} e^{-i k_{x,pp} x} + \hat{\bm{y}} e^{i k_{x,pp} x}).
    \label{eqn:polarization_grating}
\end{equation}
Depending on the position along $x$, the polarization will alternate between right and left circularly polarized states, transitioning through elliptically polarized and $45\degree$ linearly polarized states, as illustrated in Fig.~\ref{fig:grating}. Alternatively, Eqn.~\ref{eqn:polarization_grating} can be reorganized in the following form\cite{cameron1996spin}:
\begin{equation}
    \bm{E}_{\rm{pp}} = E_{pp} e^{-i(\omega_{pp} t - k_{z,pp} z)} [(\hat{\bm{x}}-i \hat{\bm{y}}) \cos{ (k_{x,pp} x + \frac{\pi}{4})} e^{i\frac{\pi}{4}} + (\hat{\bm{x}}+i \hat{\bm{y}}) \sin{ (k_{x,pp} x + \frac{\pi}{4})} e^{-i\frac{\pi}{4}}].
    \label{eqn:polarization_grating_circular}
\end{equation}
Namely, the polarization grating can be treated as the superposition of a right circularly polarized beam and a left circularly polarized beam both with spatially periodic intensities but a $\pi$ spatial phase shift (Fig.~\ref{fig:grating}). Although the total intensity is spatially uniform in this case, the polarization grating can create spatially periodic excitations in materials that respond differently to right and left circularly polarized light, e.g. spintronic and valleytronic materials, which will be discussed in detail in Section~\ref{sec:spin_gratings}.

\subsection{Material Response}
The optical grating formed by the two pump beams can induce a material response through either resonant (absorption) or nonresonant processes. In a resonant process, the incident pump light is absorbed by the material, first leading to electronic excitations. The excited electrons swiftly thermalize and cool down to the band edges through electron-electron and electron-phonon interactions. This process occurs typically on the femtosecond to picosecond time scale. Before the excited charge carriers recombine (typically on picosecond to nanosecond scale), a periodic distribution of the charge carrier density (``carrier density grating'') exists in the material, the detection of which can provide information on photocarrier diffusion and recombination. During the photocarrier cooling process, energy is transferred from the charge carriers to the lattice, leading to heating and the formation of a periodic temperature profile (``thermal grating'') within the absorption region. Subsequent detection of the thermal grating can be used to monitor the thermal diffusion process in the material, which typically happens on nanosecond to microsecond time scale in most materials with common grating periods. Associated with the thermal grating is a ``stress grating'' due to thermal expansion. The stress grating can lead to surface displacements that impact the grating detection in the reflection geometry, and initiate SAWs whose wavelength matches the grating period\cite{rogers2000optical}. On the other hand, in a nonresonant process, the incident optical field directly couples to the material without absorption and changes its refractive index. This process is also termed impulsive stimulated Brillouin scattering (ISBS). The coupling between a polarization grating with a chiral material can also generate ``spin grating''\cite{cameron1996spin} or ``valley polarization grating''\cite{mahmood2018observation}. 

\subsubsection{Thermal Grating}
Once a thermal grating, or a spatially periodic temperature distribution, is formed after the absorption of the pump beams, the subsequent dynamics of the thermal grating is governed by heat transport from the high temperature locations (``peaks'' in the thermal grating) to the low temperature locations (``valleys'' in the thermal grating). When the grating period is much longer than the mean free path of the intrinsic heat carriers (electrons in metals and phonons in semiconductors and insulators), heat transport is purely diffusive and the temperature evolution is governed by the heat equation:
\begin{equation}
    \rho C_p \frac{\partial T(x,z,t)}{\partial t} =  \kappa_x \frac{\partial^2 T}{\partial x^2} + \kappa_z \frac{\partial^2 T}{\partial z^2}  + S(x,z)\delta(t),
    \label{eqn:heat_equation}
\end{equation}
where $\rho$ is the mass density, $C_p$ is the specific heat under constant pressure, $\kappa_x$ and $\kappa_z$ are the in-plane and cross-plane thermal conductivities, respectively, and $S(x,z)$ is a source term describing the heat input from the pump beams:
\begin{equation}
    S(x,z) \propto \cos(qx) e^{-\frac{z}{d_a}},
\end{equation}
where $d_a$ is the absorption depth. In most cases, the duration of the pump pulses is much shorter than the heat diffusion time, so the pump input can be treated as instantaneous and represented with a delta-function $\delta(t)$.

Equation~\ref{eqn:heat_equation} can be solved using spatial and temporal Fourier transforms\cite{johnson2011optical,robbins2019exploring,kim2020investigation} assuming the temperature change is small, which is often ensured in TGS experiments. The simplest scenario is when the sample is weakly absorbing and the sample thickness $d$ is much smaller than the absorption depth $d_a$. In this case, the heat generation is approximately uniform along the depth ($z$) direction, and the heat diffusion is purely one-dimensional along $x$. The solution is
\begin{equation}
    \tilde{T}(q,t) \propto e^{-\alpha_x q^2 t},
    \label{eqn:1D_T_solution}
\end{equation}
where $\tilde{T}(q,t)$ is the magnitude of the temperature profile with the wavevector $q$, and $\alpha_x = \frac{\kappa_x}{\rho C_p}$ is the in-plane thermal diffusivity. This result indicates that the magnitude of the thermal grating decays exponentially, with a time constant $\tau_T = \frac{1}{\alpha_x q^2}$. This simple relation allows for the measurement of in-plane thermal diffusivity of optically thin suspended thin films, weakly absorbing liquids and solids\cite{johnson2013direct,cuffe2015reconstructing,robbins2019ballistic}. 

If the sample is opaque ($d \gg d_a$), the pump light is absorbed near the sample surface and the heat diffusion along the $z$ direction needs to be considered. In this case, Eqn.~\ref{eqn:heat_equation} can still be solved using Fourier transforms\cite{johnson2011optical}, but a numerical integration needs to be conducted for an inverse Fourier transform to obtain the solution in real time. In the limit of surface absorption ($d_a \rightarrow 0)$, the surface temperature has the following analytical form\cite{kading1995transient}:
\begin{equation}
    \tilde{T}(q,z=0,t)=A(\alpha_z t)^{-\frac{1}{2}} e^{-\alpha_x q^2 t},
    \label{eqn:surface_T_response}
\end{equation}
where $A$ is a constant and $\alpha_z$ is the cross-plane thermal diffusivity. This functional form indicates that, although cross-plane heat diffusion contributes to the temperature response, it cannot be measured using TGS at the surface absorption limit since $\alpha_z$ cannot be distinguished from the proportional constant $A$ in the fitting process. When the absorption depth $d_a$ is finite, the $t \rightarrow 0$ response will not diverge and the short-$t$ response can, in principle, be used to extract $\alpha_z$. In practice, however, the short-$t$ response may contain other contributions, e.g. from photocarrier diffusion, so is typically not used to analyze thermal transport. Another insight from Eqn.~\ref{eqn:surface_T_response} is that the longer-$t$ response will be dominated by the $t^{-\frac{1}{2}}$ decay and becomes less sensitive to in-plane heat diffusion. 

The solutions discussed above can be generalized to more complicated sample geometries. For example, K\"{a}ding et al. derived the temperature response at the surface absorption limit in a multi-layer structure using a transfer matrix approach\cite{kading1995transient}. Robbins derived the temperature response of a thin film on a substrate with a finite absorption depth\cite{robbins2019exploring}. Both solutions are complicated and involve numerical integrations. Due to the decreased measurement sensitivity, in general, when the number of parameters is increased in complicated samples, TGS has not been broadly used to probe thermal transport in these structures beyond the effective medium approximation, where an effective thermal diffusivity of a complicated sample is considered\cite{luckyanova2013anisotropy}. For strongly absorbing samples grown on substrates, a rule of thumb to determine whether the substrate contributes to the measured response is to evaluate the thermal diffusion length within the thermal decay time $\tau_T$: $d_T \sim \sqrt{4 \alpha_z \tau_T} = 2 \sqrt{\frac{\alpha_z}{\alpha_x q^2}} = \sqrt{\frac{\alpha_z}{\alpha_x}} \frac{L_S}{\pi}$. For an isotropic material, $d_T \sim \frac{L_S}{\pi}$. Therefore, when the sample thickness is larger than the optical grating period, heat does not penetrate into the substrate within the effective measurement time window and the substrate contribution can be neglected.

\subsubsection{Stress Grating and Surface Acoustic Wave}
Associated with the induced temperature grating is a stress grating due to thermal expansion. The periodic distribution of the thermal stress can lead to two effects that can be detected in TGS: a static surface displacement directly associated with the surface temperature distribution and a dynamic elastic response, i.e. SAWs, launched by the initial thermal stress. The governing equation of the thermoelastic response in an isotropic, elastic and homogeneous material is\cite{nowacki1986thermoelasticity,duggal1992real}
\begin{equation}
    c_{44} \nabla^2 \bm{u} + (c_{11}-c_{44}) \nabla (\nabla \cdot \bm{u}) = \gamma \nabla T + \rho \frac{\partial^2 \bm{u}}{\partial t^2},
    \label{eqn:thermoelasticity}
\end{equation}
where $\bm{u} = (u_x,u_z)$ is the displacement vector, $c_{11}$ and $c_{44}$ are elastic constants related to the bulk longitudinal and transverse speeds of sound $v_L$ and $v_T$: $c_{11} = \rho v_L^2$ and $c_{44} = \rho v_T^2$, $\gamma$ is the thermal stress constant related to the elastic constants and the linear thermal expansion coefficient $\alpha_{th}$: $\gamma = (3 c_{11}-4 c_{44}) \alpha_{th}$. The static surface displacement caused by the thermal stress can be solved using Eqn.~\ref{eqn:thermoelasticity} without the dynamic term $\rho \frac{\partial^2 \bm{u}}{\partial t^2}$. Assuming surface absorption of the pump light ($d_a \rightarrow 0$), the out-of-plane displacement $u_z$ at the surface has the following analytical form\cite{kading1995transient}:
\begin{equation}
    \tilde{u}_z (q,z=0,t) \propto e^{-q^2 |\alpha_x - \alpha_z|t} \mathrm{erfc} (q \sqrt{\alpha_x t}),
    \label{eqn:surface_displacement_grating}
\end{equation}
where $\mathrm{erfc}(x)=(2\pi)^{-\frac{1}{2}} \int_x^{+\infty} e^{-s^2} ds$ is the complementary error function. For a thermally isotropic material, the time dependence of the surface displacement response simply follows $\mathrm{erfc} (q \sqrt{\alpha_x t})$. As will be discussed in Section~\ref{sec:thermal_transport_applications}, the thermally induced surface displacement grating can contribute to the TGS signal in the reflection geometry and, thus, be used to extract the in-plane thermal diffusivity $\alpha_x$. The surface placement solution can also be generalized to a multi-layer structure assuming surface absorption\cite{kading1995transient}. In addition to the static surface displacement, the dynamic elastic response of the material as a consequence of the initial thermal stress distribution will lead to the generation of counter-propagating SAWs, whose wavelength matches the optical grating period. The frequency and the mode shape of the SAWs can be computed using an eigen-frequency analysis of the governing thermoelastic equation\cite{duggal1992real}. 

\subsubsection{Other Responses}
In addition to thermal and stress gratings, the optical grating formed by the pump beams can induce other material responses depending on the detection time window, the grating polarization, and the sample properties. For example, a periodic distribution of photocarrier density or spin/valley polarization density can be generated under suitable conditions. In general, the subsequent evolution of these ``transient gratings'' is determined by diffusion and local relaxation processes, which can be described by a generalized governing equation:
\begin{equation}
    \frac{\partial g}{\partial t} = D \nabla^2 g - \frac{g}{\tau_r},
    \label{eqn:generalized_diffusion}
\end{equation}
where $g$ is a generalized density (photocarrier density, spin polarization density, etc.), $D$ is a generalized diffusion coefficient, and $\tau_r$ is the relaxation/recombination time. Higher order relaxation/recombination processes, e.g. radiative recombination, Auger recombination and bimolecular recombination, can be incorporated into Eqn.~\ref{eqn:generalized_diffusion} by adding terms proportional to higher powers of $g$. Assuming first-order relaxation, the decay time constant $\tau_g$ of the periodic profile of $g$ is given by:
\begin{equation}
    \frac{1}{\tau_g} = D q^2 + \frac{1}{\tau_r}.
    \label{eqn:general_relaxation_time}
\end{equation}
Since the diffusion process depends on the optical grating wavevector $q$ while the local relaxation process does not, Eqn.~\ref{eqn:general_relaxation_time} provides a straightforward framework to distinguish diffusion and relaxation processes with TGS.

\subsection{Grating Detection}
\subsubsection{General Considerations}
In a TGS experiment, the material response discussed in the previous section is manifested in the periodic change in the sample's optical properties, which can be detected by the diffraction of a probe beam, as illustrated in Fig.~\ref{fig:schematic}. Assuming the sample has an equilibrium complex refractive index $N = n + ik$, where $n$ is the refractive index and $k$ is the extinction coefficient, the transient gratings formed in the sample can lead to a spatial modulation of these optical constants: 
\begin{equation}
    \Delta N(x,t)= \Delta n(t)\cos{(qx)}+i\Delta k(t)\cos{(qx)},
\end{equation}
where $\Delta n(t)$ and $\Delta k(t)$ are the magnitudes of the index modulation and are proportional to the magnitudes of various material responses assuming the changes are small:
\begin{equation}
    \Delta n(t)=\frac{\partial n}{\partial T}\Delta T(t) + \frac{\partial n}{\partial g}\Delta g(t) + ...
\end{equation}
\begin{equation}
    \Delta k(t)=\frac{\partial k}{\partial T}\Delta T(t) + \frac{\partial k}{\partial g}\Delta g(t) + ...
\end{equation}
Since $n$ determines the propagation speed of light inside the sample, $\Delta n$ tends to distort the phase of the probe light. Therefore, the diffracted probe intensity caused by $\Delta n$ is called the ``phase grating'' contribution. In contrast, $\Delta k$ leads to the change in absorption inside the sample and, thus, the intensity of the probe light. So the $\Delta k$ contribution to the TGS signal is called the ``amplitude grating'' contribution.

\begin{figure}[hbt!]
\centering
\includegraphics[width=\columnwidth,keepaspectratio]{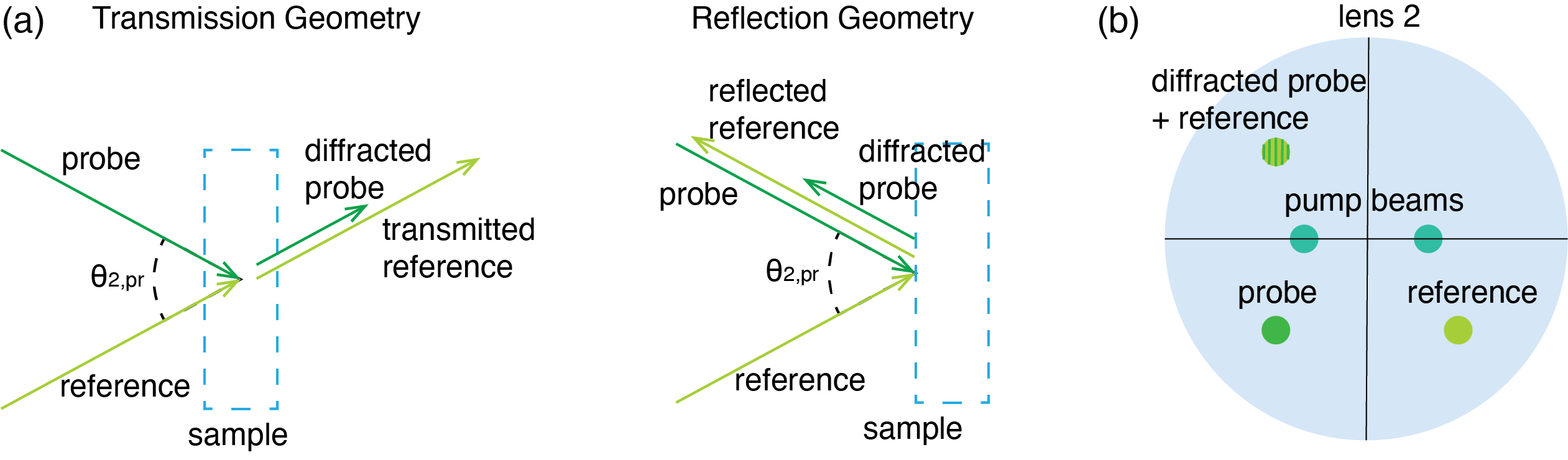}
\caption{\textbf{Schematic Illustration of the Spatial Arrangement of the Beams.} (a) [left] The beam arrangement in the transmission geometry. [right] The beam arrangement in the reflection geometry. (b) The beam arrangement on the final lens in the reflection geometry.}
\label{fig:beam_arrangement}
\end{figure}

The electric field associated with the probe beam incident onto the sample surface can be described as
\begin{equation}
    \bm{E}_{\rm{pr}}=E_{pr} \hat{\bm{p}}_{r} e^{-i (\omega_{pr} t - k_{z,pr} z + k_{x,pr} x)},
\end{equation}
where $E_{pr}$ is the field amplitude, $\hat{\bm{p}}_{r}$ is the polarization of the probe beam, $\omega_{pr}$ is the angular frequency of the probe beam, $k_{z,pr} = \frac{2 \pi}{\lambda_{pr}} \cos{\frac{\theta_{2,pr}}{2}}$ and $k_{x,pr} = \frac{2 \pi}{\lambda_{pr}} \sin{\frac{\theta_{2,pr}}{2}} \approx \pi \frac{\theta_{2,pr}}{\lambda_{pr}} = \pi \frac{\theta_{2,pp}}{\lambda_{pp}}$ (from Eqn.~\ref{eqn:pump_probe_angle}). From the diffraction theory\cite{goodman2005introduction}, the probe beam will gain an $x$-direction momentum matching the wavevector $q \approx 2 \pi \frac{\theta_{2,pp}}{\lambda_{pp}} = 2 k_{x,pr}$ of the optical grating after interacting with the transient grating formed in the material. Therefore, in a transmission geometry, a first-order diffracted probe beam has the following form:
\begin{equation}
    \bm{E}_{\rm{df,t}} \propto E_{pr} \hat{\bm{p}}_{r} e^{-i (\omega_{pr} t - k_{z,pr} z - k_{x,pr} x)}.
    \label{eqn:transmitted_probe}
\end{equation}
 Similarly, in a reflection geometry, a first-order diffracted probe beam has the following form:
\begin{equation}
    \bm{E}_{\rm{df,r}} \propto E_{pr} \hat{\bm{p}}_{r} e^{-i (\omega_{pr} t + k_{z,pr} z - k_{x,pr} x)}.
\end{equation}
The directions of the diffracted probe beams are shown in Fig.~\ref{fig:beam_arrangement}(a). As can be seen from the illustration, an important advantage of the phase-mask-based TGS setup shown in Fig.~\ref{fig:schematic} is that the diffracted probe beam is naturally collinear with the transmitted/reflected reference beam in the transmission/reflection geometry (the ``Bragg condition''), which significantly facilitates the alignment for heterodyne detection. Due to the presence of multiple beams, the spatial arrangement of the beams can be challenging, particularly in the reflection geometry and when the diffraction angle is small. To effectively manage the beam positions, typically the pump beams and the probe/reference beams are separated vertically (in the so-called ``box-car geometry''\cite{rogers1997optical}). An example arrangement of the beams on the final lens (Lens 2 in Fig.~\ref{fig:schematic}) in the reflection geometry is shown in Fig.~\ref{fig:beam_arrangement}(B). 

Both the amplitude and the phase of the diffracted probe beam can be modulated by the interaction with the transient grating formed in the material. In the transmission geometry, when the ``thin grating'' condition (the sample thickness $d \ll \frac{L_s^2}{\lambda_{pr}}$) is met\cite{eichler1986laser} and assuming the modulation of the refractive index is small, the complex transmission function $t$ is given by:
\begin{equation}
    \begin{split}
    t=\exp{(i\frac{2 \pi \Delta N d}{\lambda_{pr}})} \approx 1 + i\frac{2\pi d}{\lambda_{pr}}(\Delta n + i\Delta k)\cos{(qx)} \\
    =1 + i\frac{\pi d}{\lambda_{pr}}(\Delta n + i\Delta k) e^{iqx} +  i\frac{\pi d}{\lambda_{pr}}(\Delta n + i\Delta k) e^{-iqx}.
    \end{split}
    \label{eqn:transmission_coefficient}
\end{equation}
For thick gratings, the diffraction efficiency of the probe beam is more complicated and can be calculated using a coupled-wave approach\cite{eichler1986laser}, which is beyond the scope of this Tutorial. Combining Eqn.~\ref{eqn:transmitted_probe} and Eqn.~\ref{eqn:transmission_coefficient}, the first term corresponds to the transmitted probe beam, while the second and the third terms correspond to the diffracted probe beams. Therefore, the intensity of the diffracted probe beam is
\begin{equation}
    I_{df,t}=|\bm{E}_{df,t}|^2=E_{pr}^2 (\frac{\pi d}{\lambda_{pr}})^2 [(\Delta n)^2+(\Delta k)^2].
    \label{eqn:diffracted_intensity}
\end{equation}
From Eqn.~\ref{eqn:diffracted_intensity}, direct measurement of the intensity of the diffracted probe beam cannot distinguish the contributions from the amplitude grating ($\Delta k$) and the phase grating ($\Delta n$). In addition, $\Delta n$ and $\Delta k$ in typical TGS experiments are very small, so direct measurement of the diffracted probe beam is challenging. The solution to both issues is heterodyne detection, which we will discuss in the next section.

In the reflection geometry, in addition to the modulation of the refractive index, the periodic surface displacement caused by thermal expansion introduces an extra phase shift in the diffracted probe beam\cite{johnson2012phase}. Assuming the complex reflection coefficient $r=r'+ir''$, the diffracted probe beam in the reflection geometry has the following form:
\begin{equation}
    \bm{E}_{\rm{df,r}}=[r'+i(r''-\frac{4\pi}{\lambda_{pr}}\tilde{u}\cos{\frac{\theta_{2,pr}}{2}})] E_{pr} \hat{\bm{p}}_{r} e^{-i (\omega_{pr} t + k_{z,pr} z - k_{x,pr} x)},
\end{equation}
where $\tilde{u} = \tilde{u}(q,z=0,t)$ is the amplitude of the surface displacement grating in Eqn.~\ref{eqn:surface_displacement_grating}. The surface displacement modulates the optical path traveled by the diffracted probe beam and, thus, introduces an extra contribution to the phase grating. Similar to the transmission geometry, direct measurement of the intensity of the diffracted probe beam in the reflection geometry yields:
\begin{equation}
    I_{df,r}=|\bm{E}_{df,r}|^2=E_{pr}^2 [r'^2+(r''-\frac{4\pi}{\lambda_{pr}}\tilde{u}\cos{\frac{\theta_{2,pr}}{2}})^2].
    \label{eqn:diffracted_intensity_reflection}
\end{equation}
For thermal measurements using TGS, typically $r'$ and $r''$ are induced thermally and follow the same time dependence as the surface temperature (Eqn.~\ref{eqn:surface_T_response}), while $\tilde{u}$ follows a different time dependence (Eqn.~\ref{eqn:surface_displacement_grating}). Therefore, direct measurements of the diffracted probe intensity may reveal a mixed time dependence that is difficult to interpret. In the following section, we will discuss how heterodyne detection can solve these difficulties.

\subsubsection{Heterodyne Detection}
\label{sec:heterodyne_detection}
Heterodyne detection is a phase-sensitive detection technique~\cite{eesley1978optically,voehringer1995transient,chang1997isotropic,maznev1998optical,goodno1998ultrafast}. The general idea is to coherently mix the diffracted probe beam with a reference beam (also called a ``local oscillator'') with the same wavelength and measure the intensity of the mixed beam with a detector. In the phase-mask-based TGS implementation (Fig.~\ref{fig:schematic}), the reference beam is naturally generated when the phase mask separates the incident probe light into two beams: one is used as the probe beam, the other used as the reference beam. As discussed previously, an important advantage of the phase-mask-based implementation is that the diffracted probe beam and the transmitted/reflected reference beam are intrinsically collinear (Fig.~\ref{fig:beam_arrangement}), largely simplifying the alignment and improving the phase stability~\cite{maznev1998optical}. Since the reference beam is directly sent into the detector, typically a neutral-density (ND) filter is used in the reference beam path to reduce its intensity and avoid saturating the detector. The relative phase of the probe beam and the reference beam can be controlled by slightly tilting either the ND filter or a separate phase plate (typically a transparent plate with highly parallel surfaces) in the probe path.

In the transmission geometry, assuming the relative phase of the probe beam and the reference beam is $\Delta \phi$, the transmitted reference beam has the following form:
\begin{equation}
    \bm{E}_{\rm{ref,t}} = t_r E_{pr} \hat{\bm{p}}_{r} e^{-i (\omega_{pr} t - k_{z,pr} z - k_{x,pr} x + \Delta \phi)},
    \label{eqn:transmitted_reference}
\end{equation}
where $t_r$ is the transmission coefficient for the reference beam. Then, the interference between the diffracted probe and the transmitted reference generates an intensity:
\begin{equation}
\begin{split}
    I_{\rm{hd,t}} &= |\bm{E}_{\rm{df,t}}+\bm{E}_{\rm{ref,t}}|^2 \\
    &= \{(\frac{\pi d}{\lambda_{pr}})^2 [(\Delta n)^2 + (\Delta k)^2] + \frac{2\pi d}{\lambda_{pr}} t_r (\Delta n \sin{\Delta \phi} - \Delta k \cos{\Delta \phi}) + t_r^2 \} E_{pr}^2 \\
    & \approx [\frac{2\pi d}{\lambda_{pr}} t_r (\Delta n \sin{\Delta \phi} - \Delta k \cos{\Delta \phi}) + t_r^2 ] E_{pr}^2. 
    \label{eqn:heterodyne_transmission}
\end{split}
\end{equation}
The approximation in the last line of Eqn.~\ref{eqn:heterodyne_transmission} is due to the fact that $t_r$ is typically much larger than the diffraction efficiency of the probe beam. Thus, the time-dependent part in the measured intensity  Eqn.~\ref{eqn:heterodyne_transmission} is only 
\begin{equation}
    I_{\rm{hd,t}}(t) \propto \frac{2\pi d}{\lambda_{pr}} t_r  E_{pr}^2 [\Delta n(t) \sin{\Delta \phi} - \Delta k(t) \cos{\Delta \phi}].
    \label{eqn:heterodyne_transmission_t-dependent}
\end{equation}
As described here, the heterodyne detection has several obvious advantages. First, the contributions from the amplitude grating and the phase grating can now be separated by adjusting the relative phase $\Delta \phi$. The amplitude grating can be isolated when $\Delta \phi = 0$ or $\pi$, and the phase grating can be isolated when $\Delta \phi = \pm \frac{\pi}{2}$. Second, the desirable signal ($\Delta n$ or $\Delta k$) is amplified by the magnitude of the reference beam $t_r$. In many cases, $\Delta n$ or $\Delta k$ is typically on the order of $10^{-5}$ to $10^{-4}$. Thus, the heterodyne signal in these cases is enhanced by at least 4 to 5 orders of magnitude compared to direct detection (``homodyne detection'') that is proportional to the square of the index changes.  Third, the measured signal is now linear in the material response $\Delta n$ or $\Delta k$, thus simplifying the data interpretation. Fourth, unwanted parasitic signals that are not sensitive to $\Delta \phi$ can be easily cancelled out by conducting the measurements at different $\Delta \phi$'s and subtracting the results. The most commonly encountered parasitic signal is the scattered pump light into the detector, which can be easily removed using heterodyne detection. To simultaneously record the signals with two different heterodyne phases, Dennett and Short proposed adding an extra probe/reference pair with separate phase control\cite{dennett2017time}. 

Similarly, the heterodyne intensity in the reflection geometry can be derived, and the time-dependent part is\cite{johnson2012phase}:
\begin{equation}
    I_{\rm{hd,r}}(t) \propto t_r E_{pr}^2 \{r'(t)\cos{\Delta \phi}-[r''(t)-\frac{4 \pi}{\lambda_{pr}}\tilde{u}\cos{\frac{\theta_{2,pr}}{2}}]\sin{\Delta \phi}\}.
\end{equation}
In contrast to the transmission geometry, the surface displacement grating contributes to the phase grating signal in the reflection geometry and needs to be carefully considered when interpreting the measurement result. 

\begin{figure}[hbt!]
\centering
\includegraphics[width=0.6\columnwidth,keepaspectratio]{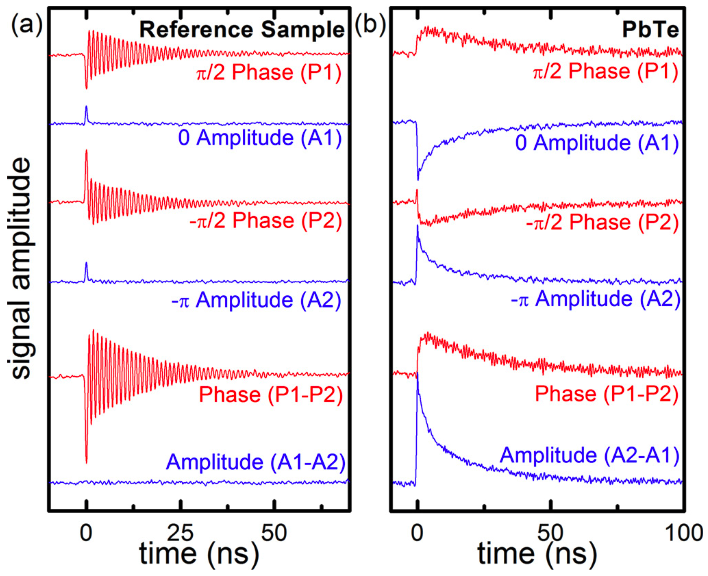}
\caption{\textbf{An Example for Heterodyne Phase Calibration.} (a) The TGS traces of a calibration sample (m-xylene) measured at different heterodyne phases. The small peak near time zero in the amplitude grating signal was due to scattered pump light into the detector. (b) The TGS traces of the sample of interest (PbTe thin film), using the heterodyne phase calibrated in (a). The amplitude grating and the phase grating signals showed different time dependence. Figure adapted from~\cite{johnson2012phase} with permission. Copyright: American Institute of Physics.}
\label{fig:heterodyne_phase}
\end{figure}

To fully exploit the above-stated advantages of heterodyne detection, it is important to calibrate the phase difference $\Delta \phi$ between the probe beam and the reference beam. One method is to use calibration samples that are known to only show phase grating or amplitude grating responses. Examples include transparent liquids~\cite{goodno1998ultrafast,johnson2012phase} and insulating polymers\cite{robbins2019exploring} that only yield a phase grating signal in the transmission geometry. One informative example by Johnson et al.\cite{johnson2012phase} is shown in Fig.~\ref{fig:heterodyne_phase}. In this case, liquid m-xylene was used as the calibration sample, whose TGS heterodyne signals were shown in Fig.~\ref{fig:heterodyne_phase}(a). Since liquid m-xylene exhibited no amplitude grating response (the small peak near time zero in the amplitude grating measurements was due to scattered pump light, which was removed by subtracting the two amplitude grating measurements), the heterodyne phase can be fixed by maximizing the phase grating signal containing acoustic oscillations. This calibrated heterodyne phase was then used to measure the sample of interest, in this case, a PbTe thin film. It can be seen from the TGS signal that PbTe showed both amplitude grating and phase grating responses with different time dependence. One issue with this calibration method, however, is that the phase difference $\Delta \phi$ in a calibration sample may differ from that in another sample. In the reflection geometry, a practical rule of thumb is that the SAW response is typically only present in the phase grating response [e.g. PbTe in Fig~\ref{fig:heterodyne_phase}(b)]. Therefore, the pure amplitude grating can be obtained by minimizing the SAW oscillations in the signal by adjusting the phase plate~\cite{johnson2012phase,dennett2018thermal}. An alternative method, proposed by Gedik and Orenstein\cite{gedik2004absolute}, utilized the fact that the probe beam and the reference beam in a phase-mask-based TGS setup are symmetrical and interchangeable, and thus, the phase difference $\Delta \phi$ can be calibrated by measuring the heterodyne intensities twice when using either beam as the probe and the other as the reference.    

\subsubsection{Practical Instrumentation Considerations}
In this section, we discuss several practical considerations in implementing a TGS setup. To detect slower processes such as thermal diffusion and SAWs, TGS setup with nanosecond time-resolution is usually sufficient. For a nanosecond setup, pulsed nanosecond laser is often used as the pump source and a CW laser is used as the probe source. For effective heterodyne detection, highly coherent single-longitudinal-mode (SLM) lasers are desirable as the probe source. To minimize steady sample heating, the CW probe laser is typically modulated with a small duty cycle to be on only within the measurement time window. The heterodyne intensity is often measured by a fast photodiode connected to an oscilloscope.  Time-resolution in this case is determined by the bandwidth of the detector and the oscilloscope, typically in the range of a few GHz. When an AC-coupled photodiode (such as Hamamatsu C5658 silicon avalanche photodiode) is used, the DC component in Eqn.~\ref{eqn:heterodyne_transmission} is automatically removed. Otherwise, usually a balanced detector is used to remove the DC component from the signal. 

To detect faster processes, such as photocarrier transport, a TGS setup with femtosecond time-resolution is required\cite{kim2017elastic}. In a femtosecond setup, both the pump and the probe beams are from an ultrafast femtosecond pulsed laser. The time resolution in this case is obtained by controlling the relative delay between the pump and the probe pulses using a mechanical delay stage. Typically the pump beam is modulated and the signal is detected by a lock-in amplifier, which also removes the DC component in the signal. Compared to the nanosecond setup, extra care needs to be taken to ensure the temporal overlap of the probe and the reference pulses when adjusting the heterodyne phase. 

Finally, we briefly comment on the choice of the pump and the probe beam sizes. So far, our analysis has assumed that both pump and probe beams are plane waves while, in reality, they are both Gaussian beams with finite beam sizes. Although smaller beam sizes lead to higher optical intensity, additional considerations should be given to optimizing the diffraction efficiency. The finite-beam-size effect on transient grating diffraction efficiency in the transmission geometry was theoretically addressed by Siegman\cite{siegman1977bragg}. Since the ratio of the angular spread of the diffracted beam to the diffraction angle is determined by the number of diffraction fringes inside the pump beam area, at least 10 diffraction fringes inside the pump beam area was recommended\cite{siegman1977bragg} to ensure the diffracted beam is well separated from the transmitted probe beam, indicating a pump beam size larger than 10 times of the grating period is desirable. Furthermore, an optimum ratio between the probe size and the pump size exists that maximizes the diffraction efficiency given a particular ratio of the pump/probe wavelengths. When the pump and probe wavelengths are similar, the optimum probe/pump size ratio is about 0.6~\cite{siegman1977bragg}. In addition, from the modeling standpoint, the beam sizes need to be large enough compared to the grating period to ensure the in-plane transport is approximately one-dimensional\cite{robbins2019exploring}. In practice, the pump and probe beam sizes need to picked to balance the optical intensity, the diffraction efficiency, and the modeling constraints. 

\section{Thermal Transport Applications}
\label{sec:thermal_transport_applications}
\subsection{General Considerations}
TGS with heterodyne detection, as described in the previous section, has been broadly used to measure the thermal diffusivity of various samples, including suspended membranes\cite{johnson2013direct,cuffe2015reconstructing,kim2017elastic,robbins2019ballistic}, thin films\cite{johnson2012phase,vega2019reduced} and bulk materials\cite{johnson2015non,dennett2018thermal,li2021remarkably}. In the transmission geometry used for measuring suspended membranes and optically thin sample ($d \ll d_a$), the temperature is uniform along the thickness direction and the temperature grating profile decays exponentially (Eqn.~\ref{eqn:1D_T_solution}). Since both $\Delta n$ and $\Delta k$ in Eqn.~\ref{eqn:heterodyne_transmission_t-dependent} follow the time dependence of the temperature profile, the heterodyne intensity decays exponentially regardless of the heterodyne phase $\Delta \phi$. In this case, heterodyne detection is mainly for amplifying the signal and removing the  phase-insensitive background. 

\begin{figure}[hbt!]
\centering
\includegraphics[width=\columnwidth,keepaspectratio]{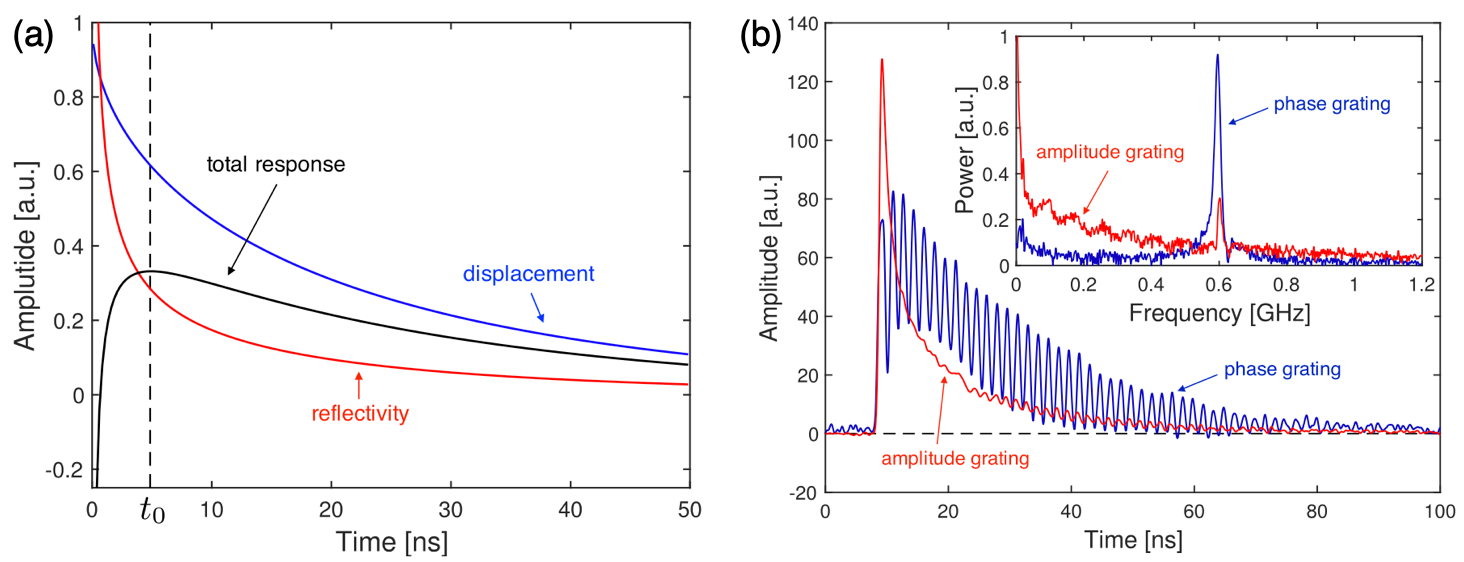}
\caption{\textbf{Measuring Thermal Diffusivity using Phase Grating Signal.} (a) Calculated phase grating signal from germanium in the reflection geometry. The total phase grating response consists of contributions from the thermally induced reflectivity change (``reflectivity'') and the surface displacement (``displacement''). (b) The experimental phase grating and amplitude grating traces of germanium. The inset shows the Fourier transport of the amplitude and phase grating signals, indicating the maximized acoustic oscillation in the phase grating signal. Figure adapted from~\cite{dennett2018thermal} with permission. Copyright: American Institute of Physics.}
\label{fig:phase_grating_thermal}
\end{figure}

In the reflection geometry for measuring opaque samples, however, care needs to be taken since the surface temperature and the surface displacement dynamics follow different time dependence (Eqn.~\ref{eqn:surface_T_response} and Eqn.~\ref{eqn:surface_displacement_grating}). For this reason, most measurements of the thermal diffusivity of opaque samples with TGS in the reflection geometry only utilize the amplitude grating signal, which isolates the contribution from the surface temperature dynamics\cite{johnson2012phase,huberman2017unifying}. In practice, however, the detection of the amplitude grating requires more stringent phase calibration than the phase grating\cite{dennett2018thermal}. If phase grating is used to extract the thermal diffusivity, in principle, both the surface temperature and surface displacement dynamics need to be included in the fitting model\cite{dennett2018thermal}. One example from~\cite{dennett2018thermal} is shown in Fig.~\ref{fig:phase_grating_thermal}, where the thermal diffusivity of bulk germanium was measured using TGS in the reflection geometry. Figure~\ref{fig:phase_grating_thermal}(a) illustrates a model calculation of the phase grating signal, which contains contributions from both the surface-temperature-induced reflectivity change and the surface displacement. The experimental phase grating and amplitude grating signals are shown in Fig.~\ref{fig:phase_grating_thermal}(b), where the phase grating signal also included acoustic oscillations. In certain samples whose reflectivity is insensitive to temperature, the phase grating signal is dominated by the surface displacement dynamics and the thermal diffusivity can be extracted by fitting the phase grating signal with Eqn.~\ref{eqn:surface_displacement_grating} alone\cite{huberman2019observation,hofmann2015non,vega2019reduced}. Another complication of using the phase grating signal is the presence of the acoustic oscillations. To accurately extract the thermal diffusivity, either the acoustic oscillations need to be removed from the signal using Fourier-transform-based techniques or a decaying sinusoidal function needs to be added to the fitting model\cite{dennett2018thermal}. 

\subsection{Ballistic Phonon Transport and Phonon Mean Free Path Spectroscopy}
A unique feature of TGS is the ease to adjust the probed length scale of thermal transport across a wide range, which is set by the grating period. When the transport length scale becomes comparable or smaller than the intrinsic mean free path (MFP) of heat-carrying phonons, heat diffusion is no longer diffusive and Fourier's law will break down~\cite{chen2021non}. Instead, the phonon Boltzmann transport equation (BTE) needs to be solved to accurately describe the thermal transport. This so-called ``quasi-ballistic regime'' of heat transport has attracted intense research interest in the past two decades due to its significant implications in applications including microelectronics thermal management~\cite{schleeh2015phonon} and nanostructured thermoelectric materials\cite{liao2015nanocomposites}. Comprehensive reviews of ballistic phonon transport can be found elsewhere\cite{hua2020phonon}. Here, we briefly review the application of TGS to probe quasi-ballistic phonon transport. We note that the TGS measurements were in parallel with and complementary to developments in detecting quasi-ballistic phonon transport with other methods\cite{minnich2011thermal,regner2013broadband,wilson2014anisotropic,siemens2010quasi}.

Effective observation of non-diffusive transport at micro/nano distances requires an experimental configuration that can be quantitatively compared to theoretical models. Heat transfer models preferred by theoreticians typically involve transport through a slab of material sandwiched between two blackbody walls with ideal thermal contacts at the interfaces~\cite{chen1998thermal}. Reproducing this sort of configuration is extremely difficult as perfect phonon blackbodies do not exist, and any interface between two real materials will be imperfect and result in thermal boundary resistance. Laser-induced TGS offers an experimental pathway to observe thermal transport in a simple geometry while also allowing for a rigorous theoretical analysis of the system. Being a non-contact optical experiment, this scheme does not involve heat transfer across any interfaces, and the excitation length scale can be easily controlled by varying the period of the interference pattern.  Additionally, the sinusoidal interference pattern results in quasi 1D transport in optically thin samples and facilitates rigorous, relatively simple theoretical treatments.   

The idea of observing nondifffusive phonon transport via TGS was first discussed theoretically by Maznev et al. in 2011~\cite{maznev2011onset}. They modeled the decay of the temperature grating using a two-population model, where phonons with a MFP much shorter than the relevant transport distance (the grating period) were characterized by a thermal diffusion model whereas long MFP phonons were modeled by the phonon BTE. The long MFP phonons interact with the thermal reservoir of short MFP phonons, but not with one another. Within this model, they found that the contribution to thermal conductivity from long MFP phonons is suppressed at smaller grating periods comparable to their MFP and the effective thermal conductivity measured by TGS at small grating periods will deviate from the bulk value, a signature for the onset of quasi-ballistic phonon transport. Assuming the material is isotropic and the relaxation time approximation is valid, the authors concluded that this suppression effect can be quantified by a simple ``suppression function'':
\begin{equation}
    A(q\Lambda) = \frac{3}{q^2 \Lambda^2}[1-\frac{\arctan{(q\Lambda)}}{q\Lambda}],
\end{equation}
where $\Lambda$ is the MFP of a specific phonon mode.
Based on this model, it was predicted that a significant reduction of the effective thermal conductivity of silicon at room temperature can be measured with TGS for grating periods as large as 10 $\mu m$. This calculation was further refined by Collins et al. by solving the BTE for all phonons and considering the spectral dependence of phonon scattering rates~\cite{collins2013non}. Monte Carlo simulation~\cite{zeng2016monte} and a variational method~\cite{chiloyan2016variational} were also used to incorporate the phonon-boundary scatterings in thin suspended membranes. These analyses were largely facilitated by the 1D periodic geometry of TGS that only requires the solution of the governing equation with a single Fourier component.  

\begin{figure}[hbt!]
\centering
\includegraphics[width=\columnwidth,keepaspectratio]{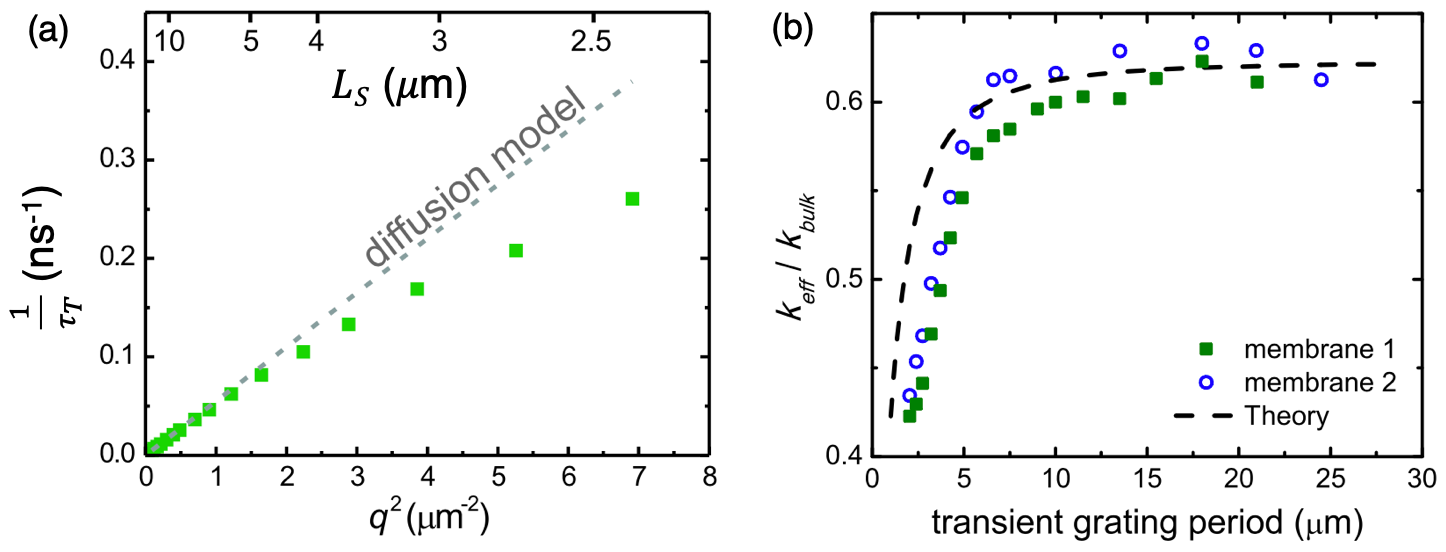}
\caption{\textbf{Quasi-ballistic Phonon Transport in Silicon Membrane Detected by TGS at Room Temperature.} (a) Measured thermal decay rate (green squaers) as a function of the grating wavevector. The deviation from the $q^2$ dependence (``diffusion model'') signals the onset of quasi-ballistic phonon transport. (b) The extracted effective thermal conductivity of the silicon membranes as a function of the grating period, showing a suppression of the thermal conductivity from the bulk value by quasi-ballistic phonon transport at small grating periods. Figure adapted from~\cite{johnson2013direct} with permission. Copyright: American Physical Society.}
\label{fig:ballistic_phonon}
\end{figure}

The theoretical prediction was verified by Johnson et al., who reported experimental observation of non-diffusive thermal transport in free-standing silicon membranes\cite{johnson2013direct} using TGS in the transmission geometry. Their representative results are shown in Fig.~\ref{fig:ballistic_phonon}. They fabricated the membranes by backside etching of a silicon-on-oxide wafer, ultimately producing 400 $nm$ thick membranes. They collected data at $\sim15$ different grating periods ranging from 2.4 to 25 $\mu m$.  Typical wave-forms featured sharp negative peaks corresponding to the electronic excitation and transport, followed by a slowly decaying thermal signal. The order of magnitude difference in the ambipolar carrier diffusion coefficient and thermal diffusivity in silicon ensures that the photocarrier grating and thermal grating are well separated in time.
Because they measured a suspended membrane, the temperature grating profile should decay exponentially, following Eqn.~\ref{eqn:1D_T_solution}.
They found that the thermal decay remains exponential for all of the grating periods considered, but the decay rate deviates from the expected $q^2$ trend as the grating period decreases [Fig.~\ref{fig:ballistic_phonon}(a)]. This departure from the $q^2$  trend signals the emergence of non-diffusive phonon transport. At large grating periods, phonons initialized in the intensity grating peaks transport to nulls, relaxing the grating and manifesting as diffusive transport. As the grating period decreases, some low frequency acoustic phonons transport ballistically from intensity peak to peak, and do not contribute to the relaxation of the grating, laeding to a reduced effective thermal conductivity [Fig.~\ref{fig:ballistic_phonon}(b)]. In a complementary experiment\cite{cuffe2015reconstructing}, Cuffe et al. used TGS to measure the effective thermal conductivity of suspended silicon membranes with varied thicknesses down to 15 $nm$. They found that, even with large grating periods up to 21 $\mu m$, the thermal conductivity of thin silicon membranes is reduced from the bulk value due to phonon scatterings by the membrane surfaces. From the thickness-dependent TGS data, they were able to reconstruct the phonon MFP distribution in silicon using an convex optimization procedure pioneered by Minnich~\cite{minnich2012determining}. More careful considerations of the phonon-boundary scattering even allowed for the extraction of phonon-wavelength-dependent specularity parameters from the TGS measurement of silicon membranes~\cite{ravichandran2018spectrally}. Additional scattering of phonons by nanoscale holes in a silicon membrane was also investigated with TGS~\cite{duncan2020thermal}. A subsequent experiment by Johnson et al.~\cite{johnson2015non} also demonstrated non-diffusive phonon transport in a bulk GaAs sample using TGS in the reflection geometry.

Recently, TGS has also been used to probe non-diffusive phonon transport in partially ordered and amorphous materials. Robbins et al. used TGS to measure the thermal conductivity of semi-crystalline polyethylene membranes that has been stretched to draw ratios up to 7.5 as a function of the grating period\cite{robbins2019ballistic}. The stretching serves to increase crystallite size and polymer chain alignment in the material, increasing thermal conductivity due to increased conduction through the stiff covalent bonds along the polymer backbone in the oriented direction. In their measurements, they found evidence of quasi-ballistic transport of phonons in the more heavily drawn samples. This manifests as a dependence of the observed thermal diffusivity on grating period down to $\sim 570$ $nm$. At smaller grating periods, the thermal diffusivity diverges from bulk values, indicating the presence of long MFP phonons that do not contribute to the relaxation of the thermal grating. Kim et al.\cite{kim2021origin} conducted TGS measurement of amorphous silicon membranes in a wide temperature range and reconstructed the phonon MFP distribution, from which they identified phonons with MFP up to micron scale even at room temperature in amorphous silicon.

While there has been success in experimentally probing phonon MFPs in materials like silicon, recent computational work suggests that thermally relevant phonons in many interesting materials have MFPs in the range of tens to hundreds of nanometers, making them difficult to access using visible light due to the optical diffraction limit. Recently, work has been done to advance the use of UV and EUV transient gratings~\cite{maznev2018generation,bohinc2019nonlinear,rouxel2021hard}. Using EUV pulses generated from a free electron laser, Bencivenga et al.\cite{bencivenga2019nanoscale} demonstrated the formation and detection of transient gratings with periods down to 28 $nm$, where extreme ballistic phonon transport was observed in silicon.

\subsection{Other Notable Thermal Transport Applications}
\begin{figure}[hbt!]
\centering
\includegraphics[width=\columnwidth,keepaspectratio]{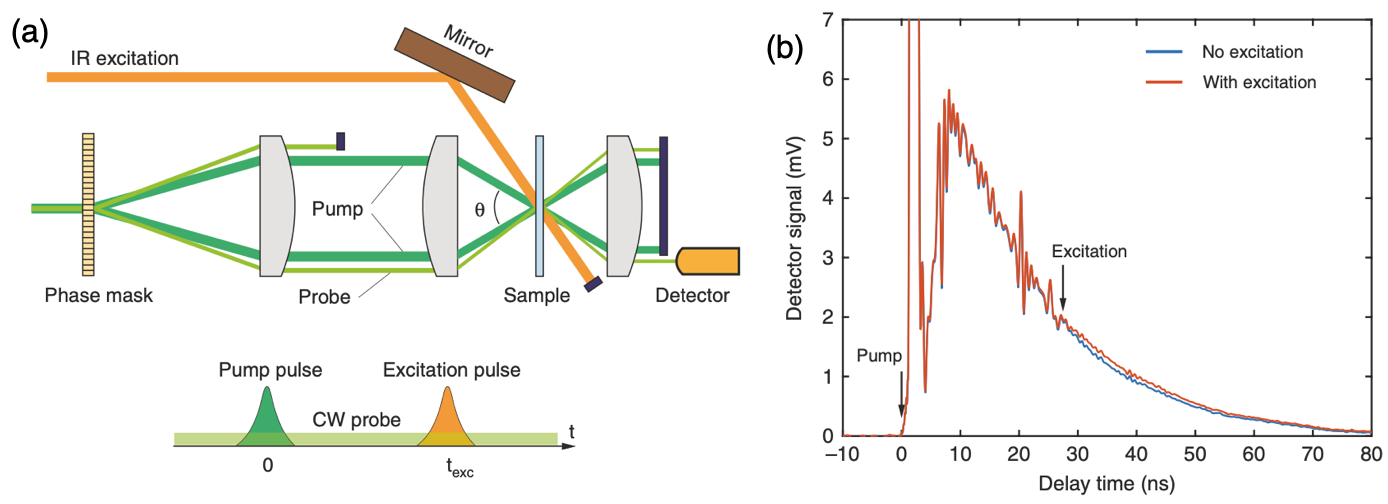}
\caption{\textbf{Impact of Electron-Phonon Interaction on Phonon Transport Quantified by TGS.} (a) The modified TGS setup to excite additional photocarriers in a silicon membrane with the Excitation beam. Thermal diffusion in the membrane is monitored after additional photocarriers are introduced. (b) The impact of the excited photocarriers on the thermal decay traces measured with TGS, showing a slowdown of the thermal decay. Figure adapted from~\cite{zhou2020direct} with permission. Copyright: Nature Publishing Group.}
\label{fig:electron_phonon_scattering}
\end{figure}

Recently, TGS has also been used to examine thermal transport phenomena that can be difficult to study with other methods. For example, Zhou et al. used a modified TGS with an additional excitation beam to directly measure the impact of electron-phonon interaction on the thermal conductivity of a silicon membrane\cite{zhou2020direct}. The scattering of phonons by free charge carriers was long thought to play a negligible role in phonon thermal transport~\cite{quan2021impact}. Recent first-principles simulation~\cite{liao2015significant} and photoacoustic measurement~\cite{liao2016photo}, however, indicated that electron-phonon scattering can significantly suppress the thermal conductivity of semiconductors even at room temperature when the electron concentration is sufficiently high. The challenge associated with experimental verification of this effect is that the free charge carrier concentration needs to be controlled while not introducing atomic defects to obscure the phonon scattering mechanism. One way to achieve this goal is to introduce charge carriers via photoexcitation, which can be naturally incorporated into photothermal measurements such as TDTR and TGS. TGS is particularly suitable for this purpose since it requires no metal transducers and a uniform distribution of free charge carriers can be most conveniently introduced in thin membranes that can be easily measured by TGS. In their TGS experiment, Zhou et al. introduced a third pulsed laser beam (the ``excitation beam'') with a wavelength of 800 $nm$ to excite electron-hole pairs with concentrations up to $7.8 \times 10^{19}$ $cm^{-3}$ in a 2-$\mu m$-thick silicon membrane [Fig.~\ref{fig:electron_phonon_scattering}(a)]. These photogenerated free charge carriers persist in the membrane within the time window of the TGS measurement and effectively reduce the measured thermal conductivity due to additional phonon scatterings by the free charge carriers[Fig.~\ref{fig:electron_phonon_scattering}(b)]. Their result indicated a nearly 30\% reduction of the thermal conductivity by electron-phonon interaction at the highest electron-hole concentration and provided direct evidence to the important role of electron-phonon interaction in thermal transport in semiconductors.

\begin{figure}[hbt!]
\centering
\includegraphics[width=\columnwidth,keepaspectratio]{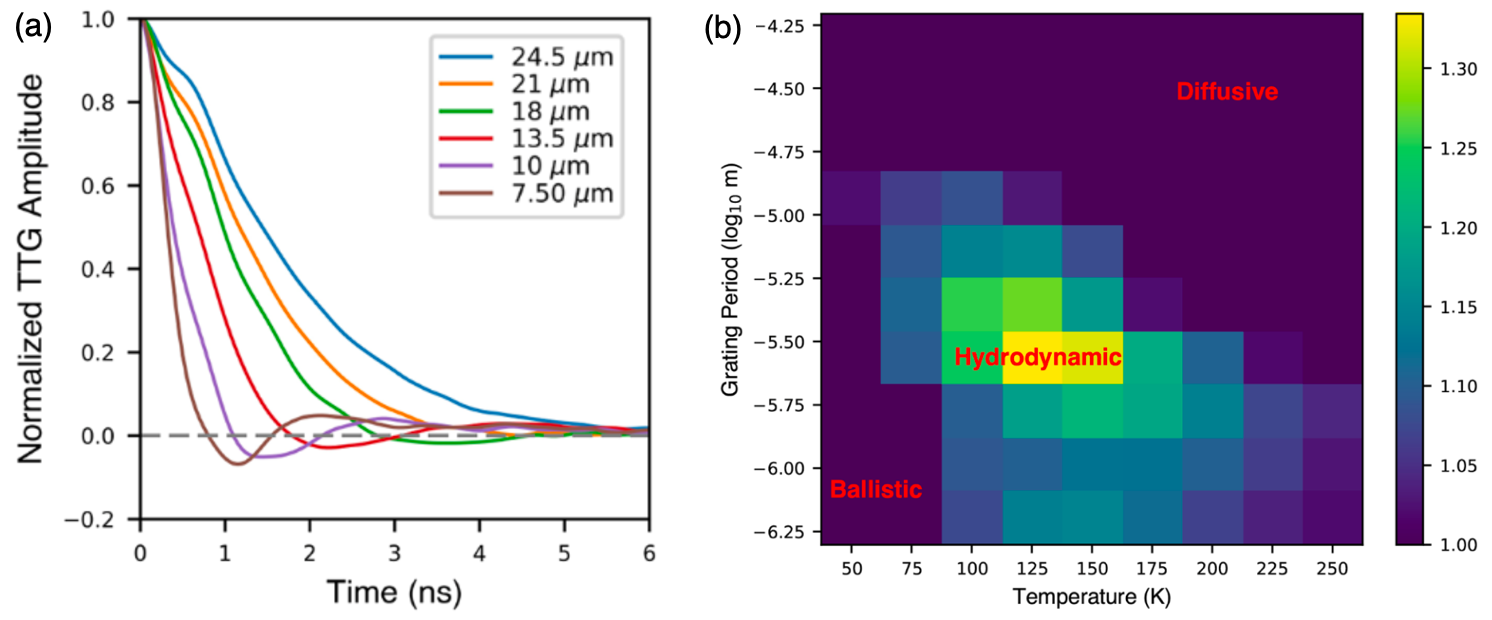}
\caption{\textbf{Observation of Second Sound in Graphite above 100 $K$ with TGS.} (a) The experimental TGS traces of graphite with a range of grating periods at 85 $K$. The negative value of the TGS response signals a $\pi$ spatial phase shift of the thermal grating, a signature for the second sound. (b) A map of various phonon transport regimes in graphite determined by the grating period and the temperature. Figure adapted from~\cite{huberman2019observation} with permission. Copyright: American Association for the Advancement of Sciences.}
\label{fig:second_sound}
\end{figure}

Another example is the detection of hydrodynamic phonon transport and second sound~\cite{beck1974phonon} with TGS~\cite{huberman2019observation}. Hydrodynamic phonon transport occurs when the momentum-conserving normal phonon-phonon scattering dominates the momentum-destroying Umklapp scattering\cite{guyer1966solution,guyer1966thermal,lee2019hydrodynamic}. In this regime, phonon transport becomes analogous to momentum-conserving molecular flows in fluids and wavelike heat conduction emerges. One signature of hydrodynamic phonon transport is the propagation of temperature waves called the ``second sound''. Earlier investigations have identified the existence of second sound in a small number of materials at cryogenic temperatures~\cite{ackerman1966second,narayanamurti1972observation,jackson1970second}. Notably, an early report of second sound measurement in NaF around 17 $K$ by Pohl and Irniger~\cite{pohl1976observation} relied on resonant scattering of a probe laser beam by a modulated transient thermal grating that matched the frequency and the wavelength of the second sound in NaF. Recent first-principles simulations predicted that hydrodynamic phonon transport can occur in two-dimensional (2D) materials~\cite{lee2015hydrodynamic,cepellotti2015phonon} and layered materials~\cite{ding2018phonon} with weak interlayer interactions within a much wider temperature range, mainly due to the strong normal scattering caused by flexural phonon modes. Using TGS measurement of graphite in the reflection geometry, Huberman et al. revealed the existence of second sound in graphite above 100 $K$~\cite{huberman2019observation}. They measured the phase grating signal from graphite, which was dominated by the surface displacement dynamics. Due to the extremely anisotropic heat transport in graphite, the TGS measurement effectively detected 1D in-plane heat transport, leading to exponential decaying phase grating signals at room temperature. When temperature was lowered to 85 $K$, however, the heterodyned TGS signal qualitatively deviated from exponential decay and even exhibited a sign reversal, as shown in Fig.~\ref{fig:second_sound}(a). In a heterodyned TGS experiment, the sign reversal signals the spatial phase of the thermal grating is shifted by $\pi$, meaning the ``peaks'' and ``nulls'' in the initial temperature profile have exchanged their positions. This cannot happen in the diffusive regime as heat always flows from hot to cold. Thus, the sign reversal in the TGS signal is a signature for the emergence of a thermal standing wave, or counter-propagating second sound waves. The authors further determined a regime map of phonon transport in graphite based on the TGS measurement [Fig.~\ref{fig:second_sound}(b)]. This experiment suggests that hydrodynamic phonon transport can contribute to thermal transport in 2D and layered materials in a wide range of temperatures and showcases the capability of TGS to probe different regimes of thermal transport ranging from diffusive to quasi-ballistic and hydrodynamic transport.   

\section{Other Applications}
\label{sec:other_applications}
Besides thermal transport, TGS has been applied to probe a wide range of energy transport processes in materials. In this section, we briefly review notable applications of TGS to probe photocarrier dynamics, spin dynamics and SAWs.

\subsection{Photocarrier Dynamics}
Following Eqns.~\ref{eqn:generalized_diffusion} and \ref{eqn:general_relaxation_time}, TGS can be used to study the diffusion and relaxation processes of photogenerated charge carriers (``photocarriers''). Earlier TGS investigations of photocarrier diffusion in a variety of semiconductors can be found in \cite{eichler1986laser}. In inorganic semiconductors, the diffusion of photocarriers is typically much faster than thermal diffusion. For example, the ambipolar diffusivity in silicon is about 2,000 $mm^2/s$ compared to the thermal diffusivity of 80 $mm^2/s$. Thus, TGS setups with picosecond and femtosecond time resolutions are required to resolve photocarrier diffusion in these materials, and the photocarrier signal and the thermal signal are often well separated in time. On the other hand, in organic semiconductors, photocarrier dynamics can occur on a similar time scale as thermal transport, where both processes need to be considered when interpreting the TGS result~\cite{ouyang2020transient}. Another consideration is the transport length scale. If the photocarrier diffusion length $l_d = \sqrt{D \tau_r}$ is much shorter than the grating period $L_S$, the photocarrier diffusion cannot be effectively detected by TGS before the photocarriers recombine. 

\begin{figure}[hbt!]
\centering
\includegraphics[width=\columnwidth,keepaspectratio]{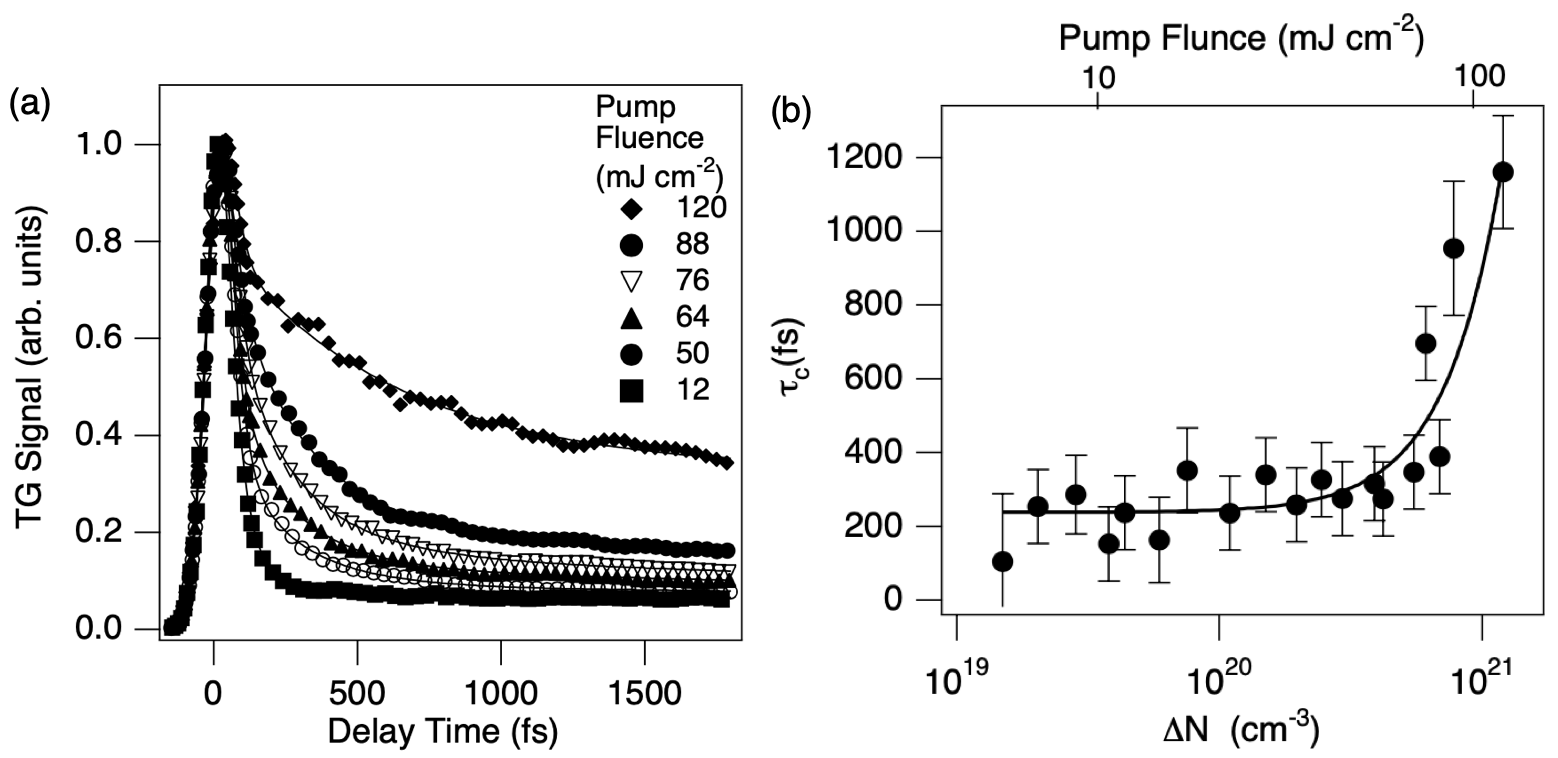}
\caption{\textbf{Photocarrier Dynamics in Silicon Probed by Femtosecond TGS.} (a) The experimental femtosecond-resolution TGS traces of silicon measured with a range of pump fluences, corresponding to photocarrier densities up to $10^{21}$ $cm^{-3}$. (b) The extracted photocarrier cooling time due to electron-phonon coupling as a function of the photocarrier density, indicating a reduced electron-phonon coupling at higher photocarrier densities due to screening. Figure adapted from~\cite{sjodin1998ultrafast} with permission. Copyright: American Physical Society.}
\label{fig:photocarrier}
\end{figure}

In a series of studies\cite{li1995measuring,li1997photoexcited,sjodin1998ultrafast}, Dai et al. used TGS in both transmission and reflection geometry to systematically study hot photocarrier transport in silicon with different photocarrier concentrations up to $10^{21}$ $cm^{-3}$. Their representative results are shown in Fig.~\ref{fig:photocarrier}. Their measurement was conducted using a femtosecond TGS system, and the results were interpreted using a general diffusion-recombination model (Eqn.~\ref{eqn:generalized_diffusion}) including various recombination channels. They found that when the photocarrier concentration is below $10^{19}$ $cm^{-3}$, carrier-carrier scattering can significantly reduce the photocarrier diffusivity\cite{li1997photoexcited}. More interestingly, when the photocarrier concentration is above $10^{20}$ $cm^{-3}$, the photocarrier lifetime increases with the concentration [Fig.~\ref{fig:photocarrier}(b)], which was attributed to screening by the photocarriers that reduces electron-phonon scattering.

In another classic application of TGS~\cite{gedik2003diffusion}, Gedik et al. used TGS to probe the diffusion of photogenerated quasi-particles in a high-temperature superconductor YBa$_2$Cu$_3$O$_{6.5}$. A unique feature of this work was the utilization of the heterodyne phase calibration procedure based on exchanging the probe and reference beams proposed by Gedik and Orenstein\cite{gedik2004absolute}. The TGS result was interpreted using a generalized diffusion equation (Eqn.~\ref{eqn:generalized_diffusion}) with first-order and second-order recombinations. At higher excitation densities, the second-order recombination dominated and the TGS signal became nearly independent of the grating period. At lower excitation densities, the diffusion process dominated and the grating-period-dependent TGS signal was used to extract the diffusion constants of the quasi-particles. This work exemplifies the capability of TGS to discern diffusion and recombination processes.

Recently, organic-inorganic hybrid perovskites have attracted intense research interest for their remarkable optoelectronic properties. TGS has been utilized to measure the photocarrier diffusion and relaxation in these materials. Webber and coworkers\cite{webber2017carrier} used femtosecond TGS to measure the photocarrier diffusion length in CH$_3$NH$_3$PbI$_3$ and showed that the photocarriers can diffuse across multiple grain boundaries before recombination. Subsequent studies\cite{scajev2017two,wang2017concentration,scajev2018diffusion} have used TGS to explore the impact of sample quality, carrier concentration and additives on photocarrier diffusion in hybrid perovskites. Arias et al.\cite{arias2018direct} further revealed the barrier effect of grain boundaries on photocarrier transport in CH$_3$NH$_3$PbI$_3$ using small grating periods comparable to the grain size. These studies showcased TGS as a versatile technique to probe both photocarrier transport and thermal transport\cite{li2021remarkably} in these emerging energy materials. 

\subsection{Spin Gratings}
\label{sec:spin_gratings}
As discussed in Section~\ref{sec:grating_formation}, the interference of two pump beams with orthogonal polarizations can form a ``polarization grating'' (or ``spin grating''), where the intensity is spatially uniform but the polarization is periodically modulated. In particular, the polarization grating can be decomposed into two periodic intensity gratings with right- and left-circular polarization\cite{cameron1996spin}. This feature implies that the polarization grating can induce spatially modulated responses in materials that respond differently to the two senses of circular polarization. This beam configuration can be easily implemented in TGS by adding a half waveplate to one of the pump beam paths. In this section, we discuss a few representative applications of spin gratings to probe electronic spin dynamics and valley polarization dynamics. 

\begin{figure}[hbt!]
\centering
\includegraphics[width=\columnwidth,keepaspectratio]{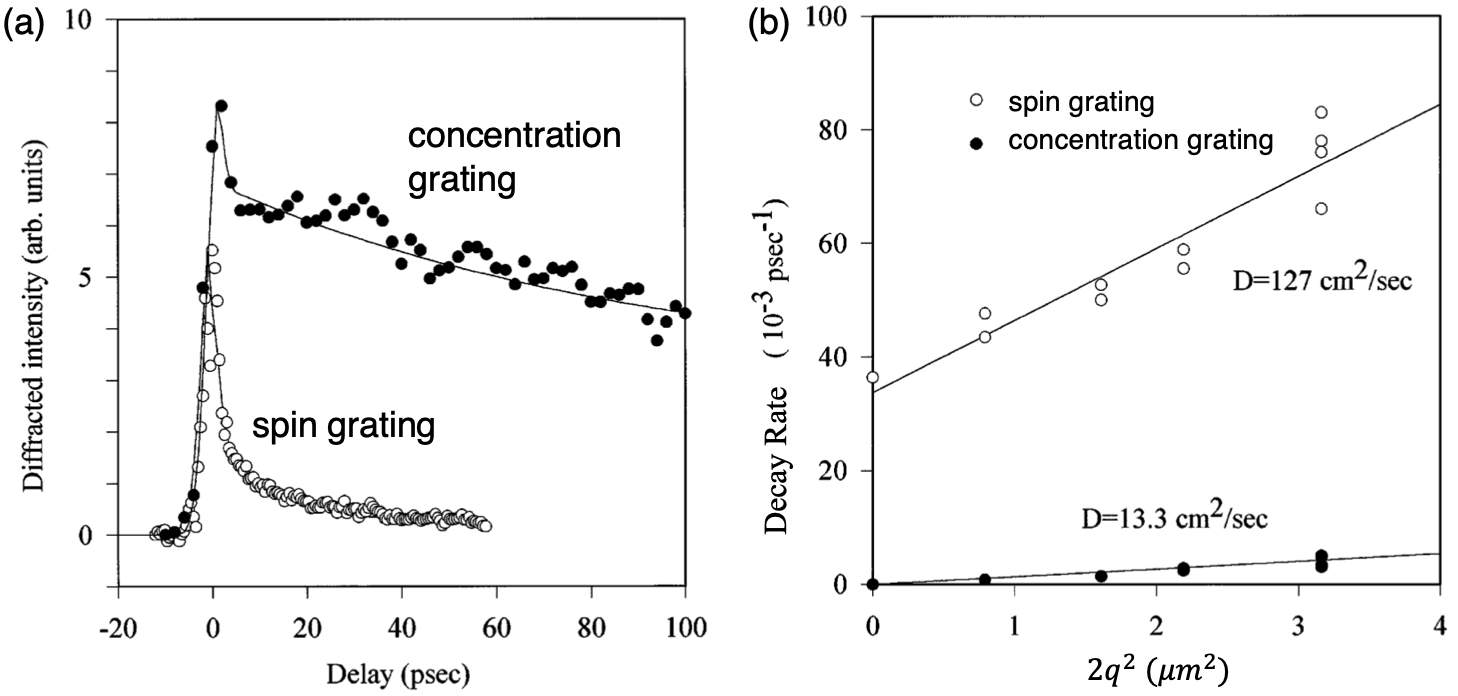}
\caption{\textbf{Transient Spin Grating Measurement of a GaAs Multiple Quantum Well.} (a) The experimental TGS traces probed by transient spin grating and concentration in the same quantum well sample. The spin grating decay is sensitive to the diffusion of electrons, which is much faster than ambipolar diffusion that is probed by the concentration grating decay. (b) The extracted electronic diffusivity and ambipolar diffusivity from the transient spin grating and concentration measurements, respectively. Figure adapted from~\cite{cameron1996spin} with permission. Copyright: American Physical Society.}
\label{fig:spin_grating}
\end{figure}

In a pioneering study, Cameron et al.\cite{cameron1996spin} utilized transient spin grating to study the electronic spin dynamics in a GaAs/AlGaAs multiple quantum well (MQW). Due to the quantum confinement effect, the split heavy hole and light hole bands in the MQW follow spin-dependent optical selection rules. As a result, circularly polarized light that resonantly excites the heavy hole band can generate pure spin-polarized electron populations in the conduction band. Excited by the optical spin grating, a transient electronic spin grating is formed that can diffract both circularly and linearly polarized probe beams, because linear polarization can be decomposed into two opposite circular polarizations. When a linearly polarized probe beam is diffracted by a transient spin grating, its polarization is rotated by 90\degree. This is in contrast to the intensity grating, where the polarization of the diffracted probe beam remains unchanged. Using two pump beams with orthogonal linear polarizations, they measured the decay of a diffracted probe beam with a linear polarization. This decay is attributed to the spin relaxation and the electronic diffusion (hole spins relax much faster and do not contribute to the spin grating decay). With this technique, they were able to separately measure the electronic diffusivity in the MQW (Fig.~\ref{fig:spin_grating}). In contrast, an intensity grating can only measure the ambipolar diffusivity of electrons and holes, which is often dominated by the hole diffusivity since holes diffuse much slower than electrons in many semiconductors. In an extension to this work, Weber et al.\cite{weber2005observation} used transient spin gratings to study electronic spin diffusion in a GaAs/AlGaAs MQW in a wide range of temperatures. At temperatures below 50 $K$, they observed ballistic transport of electronic spin, where the grating decay rate scaled linearly with the grating wavevector $q$, and oscillations in the grating decay trace were also observed. Furthermore, by comparing the spin diffusion coefficient measured using transient spin gratings to the charge diffusion coefficient from transport measurements, they concluded that the scattering mechanisms that govern spin and charge transport are different. In particular, the electron-electron scattering that conserves charge current can dissipate spin current through the so-called ``spin Coulomb drag effect''\cite{damico2001spin}. A series of subsequent studies\cite{carter2006optical,weber2007nondiffusive,weber2011measurement,yang2012doppler,wang2013gate,yang2012coherent} using the transient spin grating further refined the details of spin transport in this material system that are essential for spintronic applications. The same technique has also been used to study the relaxation of exciton spin states in colloidal quantum dots\cite{scholes2006exciton,crisp2013coherent} that are useful for a wide range of opto-electronic applications.

A recent novel application of transient spin gratings was to probe the dynamics of valley polarizations in 2D transition metal dichalcogenides\cite{mahmood2018observation}. These materials have 2D honeycomb lattices that lack the inversion symmetry, leading to two inequivalent electron ``valleys'' in the momentum space labeled by K and K'. Spin-orbit interaction further splits the valence bands such that the electrons in the two valleys can be selectively excited by right- or left-circularly polarized light\cite{xiao2012coupled}. This novel degree of freedom allows for the development of the field of ``valleytronics''\cite{vitale2018valleytronics}. One fundamental parameter for valleytronics applications is the valley depolarization time, which measures how quickly electron population selectively excited in one valley depolarizes due to various scattering and mixing mechanisms. Transient spin grating is a natural method to detect the valley depolarization since the optical spin grating can couple to the valley electrons and form a valley polarization grating. Mahmood and coworkers conducted transient spin grating measurement of a monolayer of MoSe$_2$\cite{mahmood2018observation}. They tuned the pump and the probe wavelength to match the optical band gap to ensure maximum initial valley polarization and reduce the effective temperature of the excited electrons. They measured the valley depolarization time from 300 K down to 4 K and they found the valley depolarization time depended on the excitation density, signaling a contribution from bimolecular exciton-exciton interactions. This example suggests that transient spin gratings can be a useful tool to probe materials hosting chiral excitations, such as Weyl semimetals\cite{rees2020helicity} and topological surface states\cite{wang2013observation}.   

\subsection{Surface Acoustic Waves}
The periodic stress profile induced by the transient optical grating in a material can launch bulk and surface acoustic waves that can diffract the probe beam. Since the wavelength of the launched acoustic waves match the grating period, dispersion relations of the acoustic waves can be mapped by measuring the frequencies of the oscillations in the TGS signal at a range of grating periods. This can be achieved by a Fourier analysis of the TGS signal. Depending on the stiffness of the material, the acoustic wave frequency detected by optical TGS is typically in the sub-GHz range. With EUV excitations from a free electron laser, Maznev et al. has demonstrated generation of SAWs up to 50 GHz using TGS\cite{maznev2021generation}. 

\begin{figure}[hbt!]
\centering
\includegraphics[width=\columnwidth,keepaspectratio]{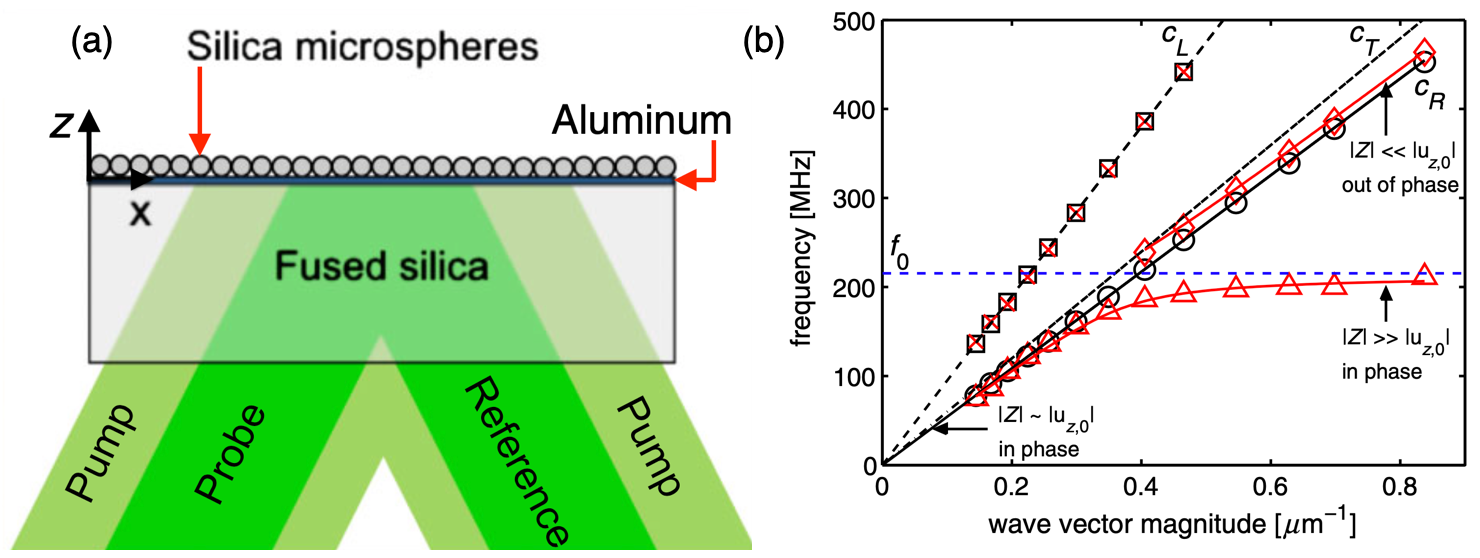}
\caption{\textbf{SAW Hybridization with Contact Resonance Detected by TGS.} (a) The sample configuration for the TGS measurement of SAW propagating near a self-assembled monolayer of silica microspheres. (b) The measured acoustic dispersions in areas without microspheres (black markers) and with microspheres (red markers). The hybridization between a SAW mode and a contact resonance mode in the area covered by microspheres leads to ``avoided crossing'' near the contact resonance frequency $f_0$. Figure adapted from~\cite{boechler2013interaction} with permission. Copyright: American Physical Society.}
\label{fig:surface_acoustic_wave}
\end{figure}

Extensive literature exists about acoustic measurements using TGS to investigate the mechanical properties of various materials\cite{eichler1986laser,rogers2000optical} and evaluate radiation damage in materials\cite{dennett2016bridging,reza2020non}. Here, we highlight recent applications of TGS to map SAW transport in self-assembled phononic crystals\cite{boechler2013interaction}. In this work, Boechler and coworkers fabricated self-assembled hexagonal array of silica microspheres on top of an aluminum coated silica substrate. Then, they conducted TGS measurement through the silica substrate and generated SAWs near the surface via absorption of the pump beams by the aluminum layer[Fig.~\ref{fig:surface_acoustic_wave}(a)]. In areas without microspheres, they were able to resolve the Rayleigh SAWs and a longitudinal acoustic wave in the silica substrate. In areas with microspheres, however, they observed a splitting of the Rayleigh SAW into two modes, which they attributed to the hybridization between the SAW and the contact resonance with the microspheres. This hybridization leads to the classic ``avoided-crossing'' behavior in the acoustic wave dispersion that effectively opens up a quasi-bandgap for acoustic waves at the crossing frequency [Fig.~\ref{fig:surface_acoustic_wave}(b)]. This phononic bandgap effect was confirmed by a following experiment by Eliason et al.\cite{eliason2016resonant} by launching and detecting SAWs at locations separated by a region covered by a self-assembled microsphere monolayer using TGS. They observed resonant attenuation of SAWs with frequency matching the avoided crossing in the dispersion relation, confirming the quasi-bandgap effect. Vega-Flick et al.\cite{vega2017vibrational} further refined the measurement and identified more acoustic modes corresponding to different contact resonance modes and the spheroidal vibrations of the microspheres. These examples demonstrate the capability of TGS to investigate acoustic wave propagation in metamaterials with complex geometrical configurations for potential applications in acoustic signal processing, filtering and medical diagnostics.  

\section{Summary and Outlook}
\label{sec:summary}
In summary, we systematically discussed the operational principles and instrumentation details of a heterodyne TGS configuration based on a diffractive phase mask. We also reviewed advancements in applying TGS to study thermal, photocarrier, spin and acoustic transport in solid-state materials. Our coverage of the literature is by no means complete and up to date, and our choices of the representative examples are mostly based on their educational values. On the outlook for future TGS developments, technical improvement will continue to push for better space and time resolutions. Although TGS with EUV sources has reached grating periods of tens of nanometers\cite{bencivenga2019nanoscale}, current implementation relies on large facilities like free electron lasers. Table-top coherent EUV and x-ray sources based on high-order harmonic generation processes\cite{rundquist1998phase,siemens2010quasi} may open up new opportunities in this realm. Access to nanometer length scale may unlock transport regimes such as extreme ballistic transport and coherent transport\cite{luckyanova2012coherent}. In the meantime, new materials and phenomena whose intrinsic transport time and length scales match current TGS capabilities can be studied. For example, energy transport in emerging chiral materials (topological Weyl semimetals and topological surface states) can be readily probe by femtosecond transient spin gratings. TGS with nanosecond and microsecond resolutions is naturally suitable for investigating dynamics in soft matter and biological materials. A recent example utilized TGS coupled with imaging to probe the photo-induced flow and phase transition in a liquid crystal\cite{katayama2019origin}. We envision that TGS will develop into a versatile material characterization platform that can probe a variety of transport processes (heat, charge, spin, and sound) in a broad range of materials (hard and soft matter) across multiple time and length scales.

\textbf{Author Declarations}

The authors have no conflicts to disclose.

\textbf{Data Availability Statement}

Data sharing is not applicable to this article as no new data were created or analyzed in this study.

\begin{acknowledgments}
This work is based on research supported by the U.S. Army Research Office under the award number W911NF-19-1-0060. Development of the TGS setup at UCSB was supported by an ARO DURIP grant under the award number W911NF-20-1-0161. Alejandro Vega-Flick also contributed to the TGS development at UCSB.
\end{acknowledgments}

\bibliography{references.bib}

\end{document}